\documentclass[aps,twocolumn,groupedaddress]{revtex4-1} 
\usepackage{graphicx}
\usepackage{amsmath}
\usepackage{amssymb}
\usepackage{import}
\usepackage{color}
\usepackage{multirow}
\usepackage{booktabs}
\usepackage{cleveref}
\crefname{equation}{Eq.}{Eqs.}  
\Crefname{equation}{Equation}{Equations}	
\crefname{figure}{Fig.}{Figs.}
\Crefname{figure}{Figure}{Figures}
\crefname{chapter}{Ch.}{Chs.}
\Crefname{chapter}{Chapter}{Chapters}
\crefname{section}{Sec.}{Secs.}
\Crefname{section}{Section}{Sections}
\crefname{appendix}{App.}{App.}
\Crefname{appendix}{Appendix}{Appendices}	
\crefname{algorithm}{Alg.}{Algs.}
\Crefname{algorithm}{Algorithm}{Algorithm}
\crefname{table}{Tbl.}{Tbls.}
\Crefname{table}{Table}{Tables}

\begin{document}
	\cmidrulewidth=.03em
	
\title{Field-induced dissociation of two-dimensional excitons in transition-metal dichalcogenides}

\author{H{\o}gni C. Kamban}
\author{Thomas G. Pedersen}

\affiliation{Department of Materials and Production, Aalborg University, DK-9220 Aalborg \O st, Denmark \\and Center for Nanostructured Graphene (CNG), DK-9220 Aalborg \O st, Denmark}

\date{\today}

\begin{abstract}
Generation of photocurrents in semiconducting materials requires dissociation of excitons into free charge carriers. While thermal agitation is sufficient to induce dissociation in most bulk materials, an additional push is required to induce efficient dissociation of the strongly bound excitons in monolayer transition-metal dichalcogenides (TMDs). Recently, static in-plane electric fields have proven to be a promising candidate. In the present paper, we introduce a numerical procedure, based on exterior complex scaling, capable of computing field-induced exciton dissociation rates for a wider range of field strengths than previously reported in literature. We present both Stark shifts and dissociation rates for excitons in various TMDs calculated within the Mott-Wannier model. Here, we find that the field induced dissociation rate is strongly dependent on the dielectric screening environment. Furthermore, applying weak-field asymptotic theory (WFAT) to the Keldysh potential, we are able to derive an analytical expression for exciton dissociation rates in the weak-field region.
\end{abstract}

\maketitle

\section{Introduction}
Interest in two-dimensional transition-metal dichalcogenide (TMD) semiconductors has increased substantially in recent years due to their exceptional electronic and optical properties. They have a wide range of applications, including photodetectors \cite{Wang2015,Lopez2013,Yin2012}, light-emitting diodes \cite{Withers2015}, solar cells \cite{Lopez2014,Bernardi2013}, and energy storage devices \cite{Guodong2010,Chhowalla2013,Soon2007}, to name a few. 
One of the most important implications of the reduced screening in two-dimensional TMDs is the comparatively large exciton binding energy \cite{Ramasubramaniam2012,Latini2015,Berkelbach2013,Pedersen2016ExcitonIonization}. Such excitons may significantly reduce the efficiency of solar cells and photodetectors, as these devices require the dissociation of excitons into free charge carriers to generate an electrical current. Excitons in bulk semiconductors will usually dissociate by thermal agitation alone due to their low binding energies. This is not the case for their two-dimensional counterparts, however, and it is therefore of great interest to obtain efficient methods of inducing exciton dissociation in TMD monolayers. Dissociation induced by in-plane static electric fields has gained attraction lately. For instance, dissociation rates for two-dimensional excitons in MoS$_2$ and $h$BN/MoS$_2$ were theoretically investigated in Ref. \cite{Haastrup2016} and for various bulk TMDs in \cite{Pedersen2016ExcitonIonization}. \\ \indent
Recently, the first systematic experimental study of field-induced dissociation of two-dimensional excitons in monolayer WSe$_2$ encapsulated by $h$BN was carried out \cite{Massicotte2018}. It was found that the limiting factor in generating photocurrents when a weak in-plane field was present was the dissociation rate of electron-hole pairs. That work also showed that the photocurrent generated in fields weaker than $15\,\mathrm{V/\mu m}$ was accurately predicted by the Mott-Wannier model \cite{Wannier1937,Lederman1976}. Nevertheless, these weak-field dissociation rates proved troublesome to obtain numerically \cite{Massicotte2018}, and they were therefore extrapolated by fitting to the rate of a two-dimensional hydrogen atom \cite{Pedersen2016Stark}.
In the present paper, we introduce a numerical method capable of computing exciton dissociation rates for significantly weaker fields with no compromise on the accuracy for stronger fields. It is based on the complex scaling approach \cite{Balslev1971,Aguilar1971} that was used in Refs. \cite{Haastrup2016} and \cite{Massicotte2018}, but, rather than rotating the entire spatial region into the complex plane, we rotate the radial coordinate only in an exterior region $r>R$. For sufficiently weak fields, we show that the rates can be obtained analytically based on the recently developed weak-field asymptotic theory (WFAT) \cite{Tolstikhin2011}, which greatly simplifies their calculation. Furthermore, we show that the weak-field ionization rate of two-dimensional hydrogen is a special case of a more general formula for dissociation of a two-dimensional two-particle system.

\section{TMD Exciton in Electrostatic Field}\label{sec:ECS}

Throughout the present paper, excitons will be modeled as electron-hole pairs described by the two-dimensional Wannier equation \cite{Wannier1937,Lederman1976}, which reads (atomic units are used throughout)
\begin{align}
\left[-\frac{1}{2\mu}\nabla^2 -w\left(\kappa\boldsymbol{r}\right)\right]\psi\left(\boldsymbol{r}\right)=E\psi\left(\boldsymbol{r}\right)\thinspace,\label{eq:UnpWan}
\end{align} 
where $\mu$ is the reduced exciton mass, $\boldsymbol{r} = \boldsymbol{r}_e-\boldsymbol{r}_h$ is the relative in-plane coordinate of the electron-hole pair, $\kappa = \left(\kappa_a+\kappa_b\right)/2$ is the average dielectric constant of the materials above and beneath the TMD sheet, and $w$ is a screened Coulomb attraction. It is well known that screening in two-dimensional semiconductors, such as TMDs, is inherently nonlocal \cite{Cudazzo2010,Keldysh1979}, i.e. momentum-dependent, and can be approximated by the linearized form $\epsilon\left(\boldsymbol{q}\right) = \kappa + r_0q$\thinspace, where $\boldsymbol{q}$ is the wave vector and the so-called screening length $r_0$ can be related to the polarizability of the sheet \cite{Cudazzo2010}. The interaction $w$ may then be obtained as the inverse Fourier transform of $2\pi\left[\epsilon\left(\boldsymbol{q}\right)q\right]^{-1}$, where $2\pi/q$ is the 2D Fourier transform of $1/r$. The resulting interaction is given by the Keldysh \cite{Keldysh1979,Trolle2017} form
\begin{align}
w\left(\boldsymbol{r}\right) = \frac{\pi}{2r_0}\left[\mathrm{H}_0\left(\frac{ r}{r_0}\right)-Y_0\left(\frac{r}{r_0}\right)\right]\thinspace,\label{eq:Keldysh_pot}
\end{align}
where $\mathrm{H}_0$ is the zeroth order Struve function and $Y_0$ is the zeroth order Bessel function of the second kind \cite{Abramowitz1974}.

When an in-plane electrostatic field is applied to the exciton, \cref{eq:UnpWan} is modified to include a perturbation term
\begin{align}
\left[-\frac{1}{2\mu}\nabla^2 -w\left(\kappa\boldsymbol{r}\right)+\boldsymbol{\varepsilon}\cdot \boldsymbol{r}\right]\psi\left(\boldsymbol{r}\right)=E\psi\left(\boldsymbol{r}\right)\thinspace.\label{eq:Wannier}
\end{align}
In the present paper, we will restrict ourselves to electric fields pointing along the $x$-axis, i.e., $\boldsymbol{\varepsilon} = \varepsilon \boldsymbol{e}_x$. As is evident, the form of \cref{eq:Wannier} is the same as that of the two-dimensional hydrogen atom in a static electric field \cite{Pedersen2016Stark}, albeit with a different potential. It should therefore come as no surprise that excitons perturbed by an electrostatic field will eventually dissociate. An important distinction, however, is that the excitons will recombine if they are not dissociated \cite{Koch2006, Massicotte2018}. This field-free recombination rate is in competition with the field-induced dissociation. For practical applications, recombination \cite{Palummo2015,Wang2016,Poellmann2015} and other forms of exciton decay (such as defect-assisted recombination \cite{Shi2013} and exciton-exciton annihilation \cite{Sun2014}) that do not yield free charge carriers, are often undesired. 

The field-induced dissociation rate $\Gamma$ is connected to the non-vanishing imaginary part of the energy eigenvalue in the presence of an electric field by the relation $\Gamma = -2\mathrm{Im}\, E$\thinspace \cite{Haastrup2016,Pedersen2016ExcitonIonization,Massicotte2018,Pedersen2016Stark}. The desired eigenvalues are therefore unobtainable through conventional Hermitian methods. Rather, one should solve \cref{eq:Wannier} subject to regularity and outgoing boundary conditions \cite{Tolstikhin2011,Siegert1936}. This is a nontrivial task in all but the simplest cases, and in practice, one usually computes the resonance energies by complex scaling the Hamiltonian \cite{Balslev1971,Aguilar1971}.

\section{Exciton Dissociation}

\begin{figure}[t]
	\resizebox{1.2\columnwidth}{!}{
\begingroup%
  \makeatletter%
  \providecommand\color[2][]{%
    \errmessage{(Inkscape) Color is used for the text in Inkscape, but the package 'color.sty' is not loaded}%
    \renewcommand\color[2][]{}%
  }%
  \providecommand\transparent[1]{%
    \errmessage{(Inkscape) Transparency is used (non-zero) for the text in Inkscape, but the package 'transparent.sty' is not loaded}%
    \renewcommand\transparent[1]{}%
  }%
  \providecommand\rotatebox[2]{#2}%
  \newcommand*\fsize{\dimexpr\f@size pt\relax}%
  \newcommand*\lineheight[1]{\fontsize{\fsize}{#1\fsize}\selectfont}%
  \ifx\svgwidth\undefined%
    \setlength{\unitlength}{312.82937179bp}%
    \ifx\svgscale\undefined%
      \relax%
    \else%
      \setlength{\unitlength}{\unitlength * \real{\svgscale}}%
    \fi%
  \else%
    \setlength{\unitlength}{\svgwidth}%
  \fi%
  \global\let\svgwidth\undefined%
  \global\let\svgscale\undefined%
  \makeatother%
  \begin{picture}(1,0.56624732)%
    \lineheight{1}%
    \setlength\tabcolsep{0pt}%
    \put(0,0){\includegraphics[width=\unitlength]{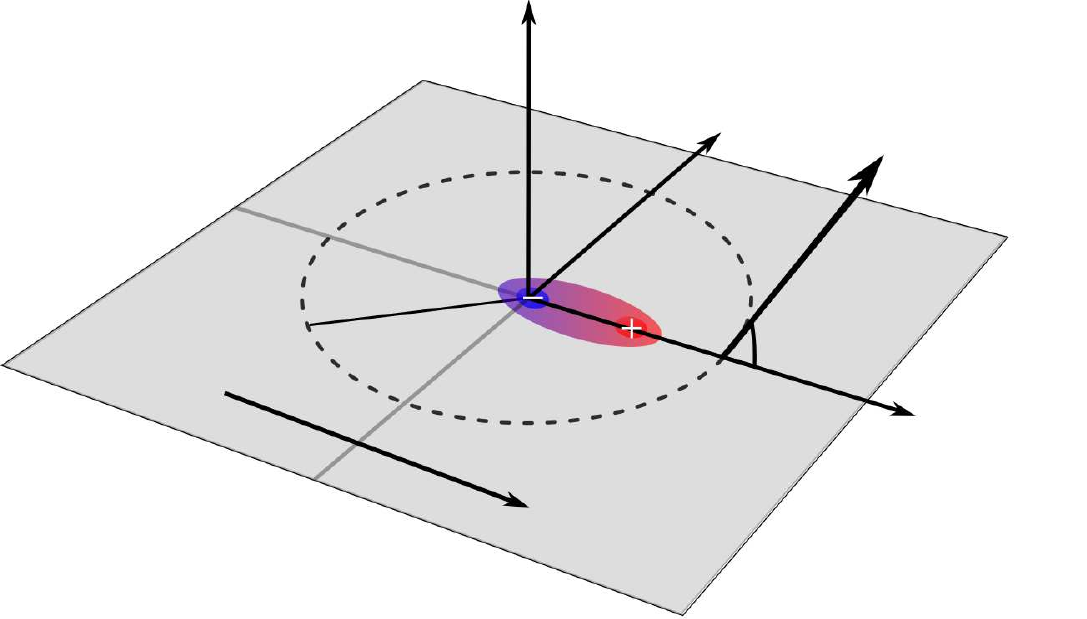}}%
    \put(0.50143208,0.53997464){\color[rgb]{0,0,0}\makebox(0,0)[lt]{\lineheight{1.25}\smash{\begin{tabular}[t]{l}$\mathrm{Im}$\end{tabular}}}}%
    \put(0.7166019,0.23453573){\color[rgb]{0,0,0}\makebox(0,0)[lt]{\lineheight{1.25}\smash{\begin{tabular}[t]{l}$\phi$\end{tabular}}}}%
    \put(0.67022976,0.44487362){\color[rgb]{0,0,0}\makebox(0,0)[lt]{\lineheight{1.25}\smash{\begin{tabular}[t]{l}$y$\end{tabular}}}}%
    \put(0.84910747,0.15203745){\color[rgb]{0,0,0}\makebox(0,0)[lt]{\lineheight{1.25}\smash{\begin{tabular}[t]{l}$x$\end{tabular}}}}%
    \put(0.54609282,0.1198513){\color[rgb]{0,0,0}\makebox(0,0)[lt]{\lineheight{1.25}\smash{\begin{tabular}[t]{l}$\mathrm{Re}$\end{tabular}}}}%
    \put(0.31708957,0.28616626){\color[rgb]{0,0,0}\makebox(0,0)[lt]{\lineheight{1.25}\smash{\begin{tabular}[t]{l}$R$\end{tabular}}}}%
    \put(0.45390063,0.07446274){\color[rgb]{0,0,0}\makebox(0,0)[lt]{\lineheight{1.25}\smash{\begin{tabular}[t]{l}$\boldsymbol{\varepsilon}$\end{tabular}}}}%
  \end{picture}%
\endgroup%
}
	\caption{Sketch of the two-dimensional exciton in the $xy$-plane with the radial coordinate rotated into the complex plane by an angle of $\phi$ for $r>R$. }\label{fig:sketch}
\end{figure}
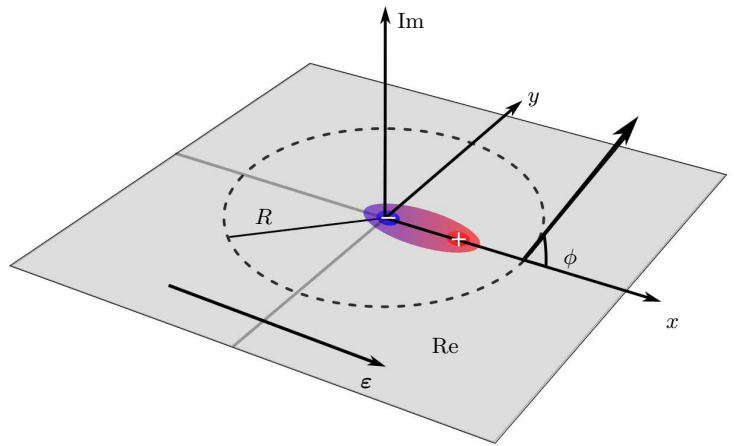

In its simplest form, complex scaling corresponds to rotating the radial coordinate into the complex plane uniformly \cite{Balslev1971,Aguilar1971} $r\to\exp\left(i\phi\right)r$\thinspace, where $\phi$ is a fixed real-valued angle (note that if $\phi$ is chosen complex the coordinate will simply be stretched as well as rotated). This transformation, referred to as uniform complex scaling (UCS), turns the outgoing waves mentioned above into exponentially decaying waves, provided that $\phi$ is chosen large enough \cite{McCurdy2004}. Thus, the complex scaled resonance wave functions are square integrable, and the resonance energies can be obtained by solving \cref{eq:Wannier} with the scaled operator and the boundary condition $\psi\left(r\to\infty\right)=0$. This approach has been used to obtain the dissociation rates of two-dimensional TMD excitons in Refs. \cite{Haastrup2016} and \cite{Massicotte2018}. Nevertheless, as was discussed briefly in Ref. \cite{Massicotte2018}, numerical difficulties arise when the electric field becomes sufficiently weak. This is because the important region for weak fields is sufficiently far from the origin that the uniformly complex scaled resonance wave function has (numerically) vanished prior to reaching this region. By utilizing the so-called exterior complex scaling (ECS) approach \cite{Simon1979,McCurdy2004,McCurdy1991,Rescigno2000}, combined with a finite element (FE) representation of the wave function, we are able to compute dissociation rates for significantly weaker fields, as we now demonstrate. 

\begin{figure}[t]
	\includegraphics[width=1\columnwidth]{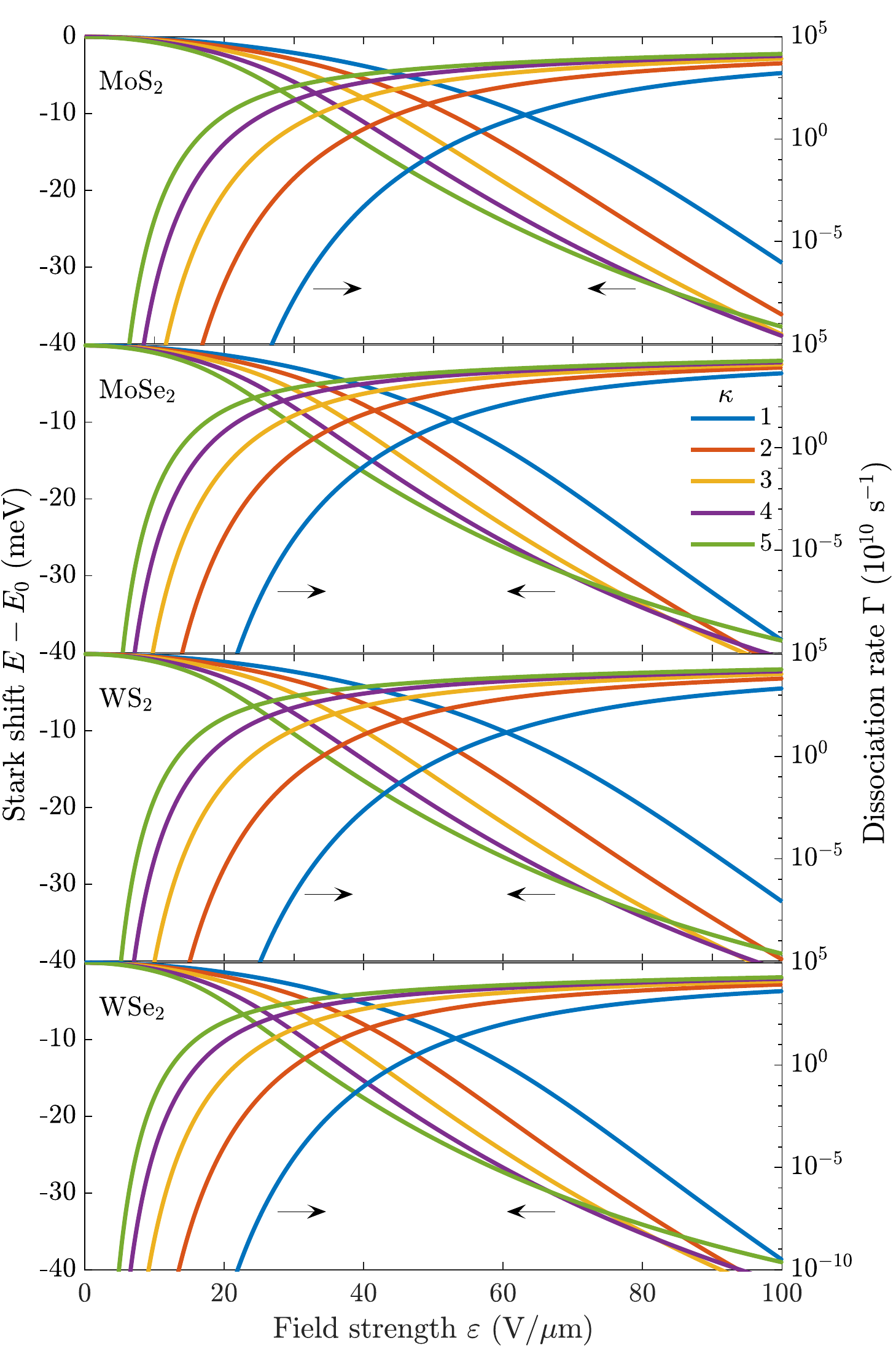}
	\caption{Exciton Stark shift and dissociation rate for four important materials in various dielectric environments. }\label{fig:stark_and_diss}
\end{figure}

As the name suggests, ECS transforms the radial coordinate outside a scaling radius $R$ 
\begin{align}
r \to \begin{cases}
r \quad &\mathrm{for }\,\, r<R\\
R + \left(r-R\right)e^{i\phi} \quad &\mathrm{for }\,\, r>R\thinspace,
\end{cases}\label{eq:ECStrans}
\end{align}
where $\phi$ is the angle of rotation, as illustrated in \cref{fig:sketch}. The partitioning of the radial coordinate is efficiently dealt with by an FE basis representation, the details of which can be found in \cref{app:numerical}. The Stark shift and dissociation rate as functions of in-plane field strength for four important materials in various dielectric environments are shown in \cref{fig:stark_and_diss}. The screening lengths and reduced masses used in the calculations are obtained from Ref. \cite{Olsen2016}. As is evident, the dissociation rate increases rapidly with increasing field strength. The rates can also be seen to be strongly dependent on the screening environment, which is to be expected as increased screening leads to reduced binding energies. It is therefore possible to tune the dissociation rates of the TMDs as desired within a certain range. For example, encapsulating the TMDs in $h$BN (with $\kappa = 4.9$ \cite{Latini2015}) increases the dissociation rates by several orders of magnitude compared to their free-space counterparts. Rates for MoS$_2$, MoS$_2$/$h$BN, and $h$BN/MoS$_2$/$h$BN were presented in Ref. \cite{Haastrup2016} for fields stronger than $50\, \mathrm{V/\mu m}$. However, the experimental study of $h$BN/WSe$_2$/$h$BN in Ref. \cite{Massicotte2018} suggests that exciton dissociation rates are the limiting factor in generation of photocurrents for applied fields weaker than $15\, \mathrm{V/\mu m}$ in this material. For stronger fields, the photocurrent measurements deviate from the field-induced rates predicted by the Wannier model, and other limitations dominate \cite{Massicotte2018}. We expect to see the same effect for the other TMDs, and we furthermore expect this threshold field to increase as the screening is reduced.\\ \indent
In weak fields, the Stark shifts in \cref{fig:stark_and_diss} can be seen to vary approximately as $\varepsilon^2$, in agreement with the lowest order perturbation theory expansion of the energy $E \approx E_0 - \frac{1}{2}\alpha\varepsilon^2$\thinspace, where $E_0$ is the unperturbed ground-state energy and $\alpha$ is the exciton polarizability. The shape of the shift is in agreement with those observed for similar systems; the energy initially decreases rapidly with field strength and then levels off as the field strength increases \cite{Pedersen2016Stark}. A more detailed analysis of the shift in the weak-field region will be made in \cref{sec:Stark}.

\section{Weak-field Asymptotic Theory}\label{sec:WFAT}
Even with our improved numerical procedure, the dissociation rates for extremely weak fields are unobtainable. In fact, any numerical procedure with finite-precision arithmetic fails for sufficiently weak fields, when the ratio $\Gamma/\left|E_0\right|$ approaches the round-off error \cite{Trinh2013,Batishchev2010}. Fortunately, with the recent development of weak-field asymptotic theory (WFAT) \cite{Tolstikhin2011}, we are able to take advantage of the simple asymptotic form of the Keldysh potential and calculate the weak-field dissociation rates analytically. To this end, we first simplify  \cref{eq:Wannier} by introducing the scaling relations
\begin{align}
\tilde{r}_0 = \frac{\mu}{\kappa^2}r_0\thinspace,\quad  \tilde{\boldsymbol{r}}=\frac{\mu}{\kappa}\boldsymbol{r}, \quad \mathrm{and} \quad \tilde{\boldsymbol{\varepsilon}} = \frac{\kappa^3}{\mu^2}\boldsymbol{\varepsilon}\thinspace,\label{eq:scaling}
\end{align}
which lead to
\begin{align}
E\left(\mu,\kappa,r_0,\varepsilon\right) = \frac{\mu}{\kappa^2}E\left(1,1,\tilde{r}_0,\tilde{\varepsilon}\right)\thinspace.\label{eq:Engscale}
\end{align}
Thus, the only nontrivial parameters are $\tilde{r}_0$ and $\tilde{\varepsilon}$\thinspace, and the analysis in the following will therefore be restricted to the simplified problem
\begin{align}
\left[-\frac{1}{2}\nabla^2 -w\left(\boldsymbol{r}\right)+\boldsymbol{\varepsilon}\cdot \boldsymbol{r}\right]\psi\left(\boldsymbol{r}\right)=E\psi\left(\boldsymbol{r}\right)\thinspace,\label{eq:SimplifiedWannier}
\end{align}
from which Stark shifts and dissociation rates can be obtained using \cref{eq:Engscale}. Note that in order to simplify the notation the tilde has been omitted in \cref{eq:SimplifiedWannier} as well as in the following. Therefore, unless explicitly stated otherwise, $r$, $r_0$, $\varepsilon$, and $E$ in the following refer to the scaled parameters. 

The potential in \cref{eq:SimplifiedWannier} has the large-$r$ behavior \cite{Abramowitz1974}
\begin{align}
w\left(\boldsymbol{r}\right) = \frac{1}{r} + O\left(\frac{r_0^2}{r^3}\right)\thinspace,\label{eq:keldyshas}
\end{align}  
which has the form required to use WFAT. A leading order expression for the weak-field dissociation rate was derived for a three-dimensional system in Ref. \cite{Tolstikhin2011} and extended to first order in $\varepsilon$ in Ref. \cite{Trinh2013}. We shall only consider the leading order approximation here. By modifying the approach in Ref. \cite{Tolstikhin2011} to two dimensions, we find that the weak-field dissociation rate for the ground-state of \cref{eq:SimplifiedWannier} is given by
\begin{align}
\Gamma \approx \left|g_0\right|^2W_0\left(\varepsilon\right)\thinspace,\label{eq:wfdiss}
\end{align}
with the asymptotic coefficient and field factor \cite{Madsen2012,Madsen2013} given by
\begin{multline}
g_0 = \lim_{v\to\infty}v^{1/2-1/k}e^{kv/2}\\
\times\int_0^\infty \varphi_0\left(u\right) \psi_0\left(\frac{u+v}{2}\right)\frac{1}{\sqrt{u}}du\label{eq:g0}
\end{multline}
and
\begin{align}
W_0\left(\varepsilon\right) = k\left(\frac{4k^2}{\varepsilon}\right)^{2/k-1/2}\exp\left(-\frac{2k^3}{3\varepsilon}\right)\thinspace,
\end{align}
respectively. Here, $k = \sqrt{-2E_0}$, and  $u$ and $v$ are the parabolic cylindrical coordinates defined by
\begin{align}
&u = r+x\thinspace,\quad u\in\left[0,\infty\right)\\
&v=r-x\thinspace,\quad v\in\left[0,\infty\right)\thinspace.\label{eq:parav}
\end{align}
The functions appearing in \cref{eq:g0} are the unperturbed ground state $\psi_0$ and 
\begin{align}
\varphi_n\left(u\right) = \left[\frac{\sqrt{k}n!}{\left(n-1/2\right)!}\right]^{1/2}L_n^{\left(-1/2\right)}\left(ku\right)e^{-ku/2}\thinspace,
\end{align}
with $L_n^{\left(\alpha\right)}\left(x\right)$ a generalized Laguerre polynomial \cite{Abramowitz1974}. To obtain the weak field dissociation rate from \cref{eq:wfdiss}, one therefore needs the unperturbed binding energy $E_0$ and the asymptotic coefficient $g_0$ of the simplified problem. Once they have been obtained, the physical weak-field dissociation rate for arbitrary monolayer TMDs can be obtained by scaling back to the original units, cf. \cref{eq:Engscale},
\begin{align}
\Gamma\left(\mu,\kappa,r_0,\varepsilon\right) = \frac{\mu}{\kappa^2}\Gamma\left(1,1,\tilde{r}_0,\tilde{\varepsilon}\right)\thinspace.\label{eq:Gammascale}
\end{align}
We now turn to computing the asymptotic coefficient $g_0$.

\subsection{Computing the asymptotic coefficient}

\begingroup
\squeezetable
\begin{table}[b]
	\caption{Binding energy $E_0\left(1,1,\tilde{r}_0\right)$ and asymptotic coefficient $g_0$ of the simplified problem for four important materials in different dielectric environments.}
	\centering
	\begin{tabular}{c cccc c cccc c cccc c cccc}
		\hline\hline\\[-0.2cm]
		\phantom &\multicolumn{2}{ c }{MoS$_2$} 
		&\phantom&\multicolumn{2}{ c }{MoSe$_2$}
		&\phantom&\multicolumn{2}{ c }{WS$_2$}
		&\phantom&\multicolumn{2}{ c }{WSe$_2$}\\
		\cmidrule{2-3}  \cmidrule{5-6} \cmidrule{8-9} \cmidrule{11-12}\\[-0.2cm]
		$\kappa$ &$E_0$ &  $g_0$ & \phantom 
		&$E_0$  & $g_0$ &\phantom
		&$E_0$ & $g_0$  &\phantom
		&$E_0$  & $g_0$  
		\\ [0.1cm]\hline\\[-0.2cm]
		$1$ & $0.0714$ & $0.00098$ &  & $0.0659$  & $0.00057$  & & $0.0921$  & $0.0044$  & & $0.0801$ & $0.0020$ \\[0.1cm]
		$2$ & $0.1907$ & $0.0889$&  & $0.1773$  & $0.0707$  & & $0.2392$  & $0.1671$  & & $0.2113$ & $0.1200$ \\[0.1cm]
		$3$ & $0.3200$& $0.3210$ & & $0.2995$  & $0.2805$   & & $0.3928$  & $0.4650$  & & $0.3512$  & $0.3829$ \\[0.1cm]
		$4$ & $0.4474$& $0.5695$ & & $0.4210$  & $0.5195$  & & $0.5392$  & $0.7323$ & & $0.4870$  & $0.6420$ \\[0.1cm]
		$5$ & $0.5680$& $0.7796$ & & $0.5370$  & $0.729$  & & $0.6740$  & $0.9380$  & & $0.6142$  & $0.8515$ \\[0.1cm]
		
	\end{tabular}\label{tbl:1}
\end{table} 
\endgroup

Finding $g_0$ given by \cref{eq:g0} requires an accurate representation of the wave function for large $v$. Note that a traditional basis expansion (e.g. a Gaussian basis) is generally not accurate enough, as only the most slowly decaying functions will contribute in this region. This problem was partially circumvented in Ref. \cite{Madsen2013} by using a Guassian basis with optimized exponents. Here, we will implement Numerov's finite difference scheme, which can accurately and efficiently construct the unperturbed wave function in the asymptotic region. The technical details can be found in \cref{app:numerov}. As a preliminary, it is convenient to relate $g_0$ to the radial wave function. The ground state of a potential with cylindrical symmetry satisfies
\begin{align}
\psi_0\left(r\right) \sim C_0r^{1/k-1/2}e^{-kr}\quad \mathrm{for}\, r\to\infty\thinspace,\label{eq:asymp}
\end{align}
where $C_0$ is a constant. Using \cref{eq:asymp} in \cref{eq:g0} leads to the relation 
\begin{align}
g_0 = \frac{2^{1/2-1/k}\pi^{1/4}C_0}{k^{1/4}}\thinspace.\label{eq:g0sym}
\end{align}
The problem of finding $g_0$ has therefore been reduced to obtaining the asymptotic coefficient of the radial wave function. It can be found by taking the limit
\begin{align}
C_0 = \lim_{r\to\infty}\psi_0r^{1/2-1/k}e^{kr}.\label{eq:C0as}
\end{align}
Note that in the unscreened limit ($r_0 \to 0$) $k=2$  and $\psi_0 = 2^{3/2}\pi^{-1/2}\exp\left[-\left(u+v\right)\right]$ \cite{Yang1991} which leads to $g_0 = 2^{5/4}\pi^{-1/4}$, and \cref{eq:wfdiss} is therefore in agreement with the expression found in Ref. \cite{Pedersen2016Stark} for the two-dimensional hydrogen atom. 
In practice, we find $C_0$ by fitting \cref{eq:C0as} to the asymptotic expansion
\begin{align}
D\left(r\right) = \sum_{n=0}^4\frac{d_n}{r^n}\thinspace,\label{eq:fitas}
\end{align}
in a stable region (see \cref{app:numerov}), as described in Ref. \cite{Madsen2013}. The asymptotic coefficient $C_0$ is then obtained by taking the limit  $\lim_{r\to\infty}D = d_0$. 
\begin{figure}[t]
	\includegraphics[width=1\columnwidth]{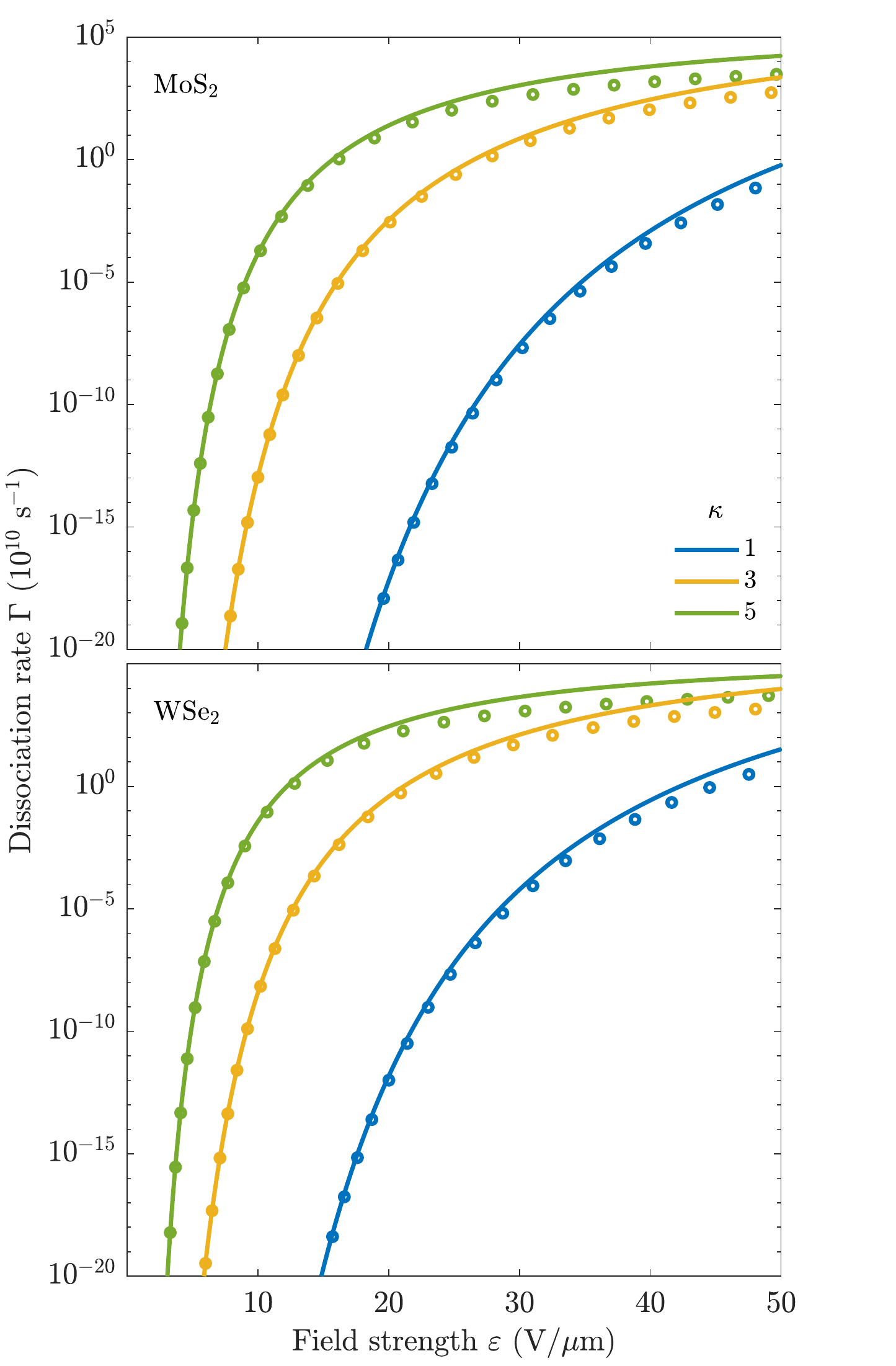}
	\caption{Exciton dissociation rates for MoS$_2$ (upper) and WSe$_2$ (lower) encapsulated by various dielectric media. The circles are the numerically exact results obtained by the method in \cref{app:numerical} (same as those in \cref{fig:stark_and_diss}). The solid lines correspond to the weak-field formula \cref{eq:wfdiss} with the parameters found in \cref{tbl:1}.}\label{fig:analytical_diss}
\end{figure}
The computational method above takes advantage of the fact that a high-order finite-difference scheme is able to accurately reproduce the wave function for large $r$. Recently, however, integral representations for the asymptotic coefficient that are insensitive to the wave function tail have been derived for a three-dimensional system \cite{Dnestryan2016,Madsen2017}. This suggests that one may get away with using a sufficiently accurate representation of the wave function only in an interior region. We shall use the integral equations as a check to ensure the accuracy of the scheme presented above. To derive the corresponding equation for our two-dimensional system we introduce the reference function $\Omega$ as a solution to
\begin{align}
\left[-\frac{1}{2}\nabla^2-\frac{1}{r}+\frac{k^2}{2}\right]\Omega_n\left(r\right)=0\thinspace.
\end{align}
The relevant function for the asymptotic coefficient of the ground state is
\begin{multline}
\Omega_0\left(r\right)=-2^{\frac{1}{k}+\frac{1}{2}} k^{\frac{1}{k}-\frac{1}{2}} \Gamma \left(\frac{1}{2}-\frac{1}{k}\right)   \\
\times\, e^{-k r} M\left(\frac{1}{2}-\frac{1}{k};1;2 k r\right)\thinspace,\label{eq:omega0}
\end{multline}
where $\, M$ is a confluent hypergeometric function \cite{Abramowitz1974}. If the exciton energy coincides with one of the energies of the two-dimensional hydrogen atom 
\begin{align}
E_n^{\left(\mathrm{hydr.}\right)} = \frac{1}{2\left(n-1/2\right)^2}\thinspace,
\end{align}
where $n=1,2,...$ \cite{Yang1991}, the confluent hypergeometric function in \cref{eq:omega0} reduces to a polynomial of finite degree and $\Omega_0$ will vanish as $r$ tends to infinity. In practical calculations, this is hardly ever the case and the reference function will therefore be exponentially increasing (see Ref. \cite{Dnestryan2018} for a discussion of the case where $E_0 \approx E_n^{\left(\mathrm{hydr.}\right)}$)
\begin{align}
\Omega\left(r\right)\sim -k^{-1}r^{-1/2-1/k}e^{kr}\quad \mathrm{for}\, r\to \infty\thinspace.\label{eq:omgas}
\end{align}
Integrating by parts and using \cref{eq:asymp,eq:omgas} when $r$ tends to infinity, we find
\begin{align}
C_0 = \int_0^\infty\Omega_0\left(r\right)\left[\frac{1}{2}\nabla^2+\frac{1}{r}-\frac{k^2}{2}\right]\psi_0\left(r\right) rdr \thinspace,
\end{align}
which, using \cref{eq:SimplifiedWannier} with $\varepsilon=0$, can be reduced to
\begin{align}
C_0 = \int_0^\infty\Omega_0\left(r\right)\left[\frac{1}{r}-w_s\left(r\right)\right]\psi_0\left(r\right) rdr \thinspace.\label{eq:Cint}
\end{align}
The integrand in \cref{eq:Cint} is a product of an exponentially increasing function $\Omega_0$ and an exponentially decreasing function $\psi_0$. Such an integral need not be convergent. Nevertheless, as is evident from the large-$r$ behavior of these functions (see \cref{eq:asymp,eq:omgas}), the exponential terms cancel for $r$ tending to infinity, resulting in the integrand tending to zero sufficiently quickly for the integral to converge. We have checked that \cref{eq:C0as,eq:Cint} agree when using the numerically exact wave function. The asymptotic coefficients and binding energies of the simplified Wannier problem for four important materials are presented in \cref{tbl:1}. Note that these binding energies increase with $\kappa$. This is because the binding energies of the simplified problem increase when $\tilde{r}_0$ decreases and $\tilde{r}_0$ is proportional to $\kappa^{-2}$. In \cref{fig:analytical_diss}, we compare the dissociation rates for excitons in MoS$_2$ and WSe$_2$ given by the weak-field formula \cref{eq:wfdiss} to the numerically exact dissociation rates. As can be seen, the agreement between the weak-field and the fully numerical results is reasonable for fields lower than $50 \mathrm{V/\mu m}$ and improves as the field strength decreases. For $\varepsilon\lesssim \kappa^{-1/2}\, 20\, \mathrm{V/\mu m}$ the agreement in \cref{fig:analytical_diss} becomes excellent.

\section{Stark Shift}\label{sec:Stark}
\begin{figure}[t]
	\includegraphics[width=1\columnwidth]{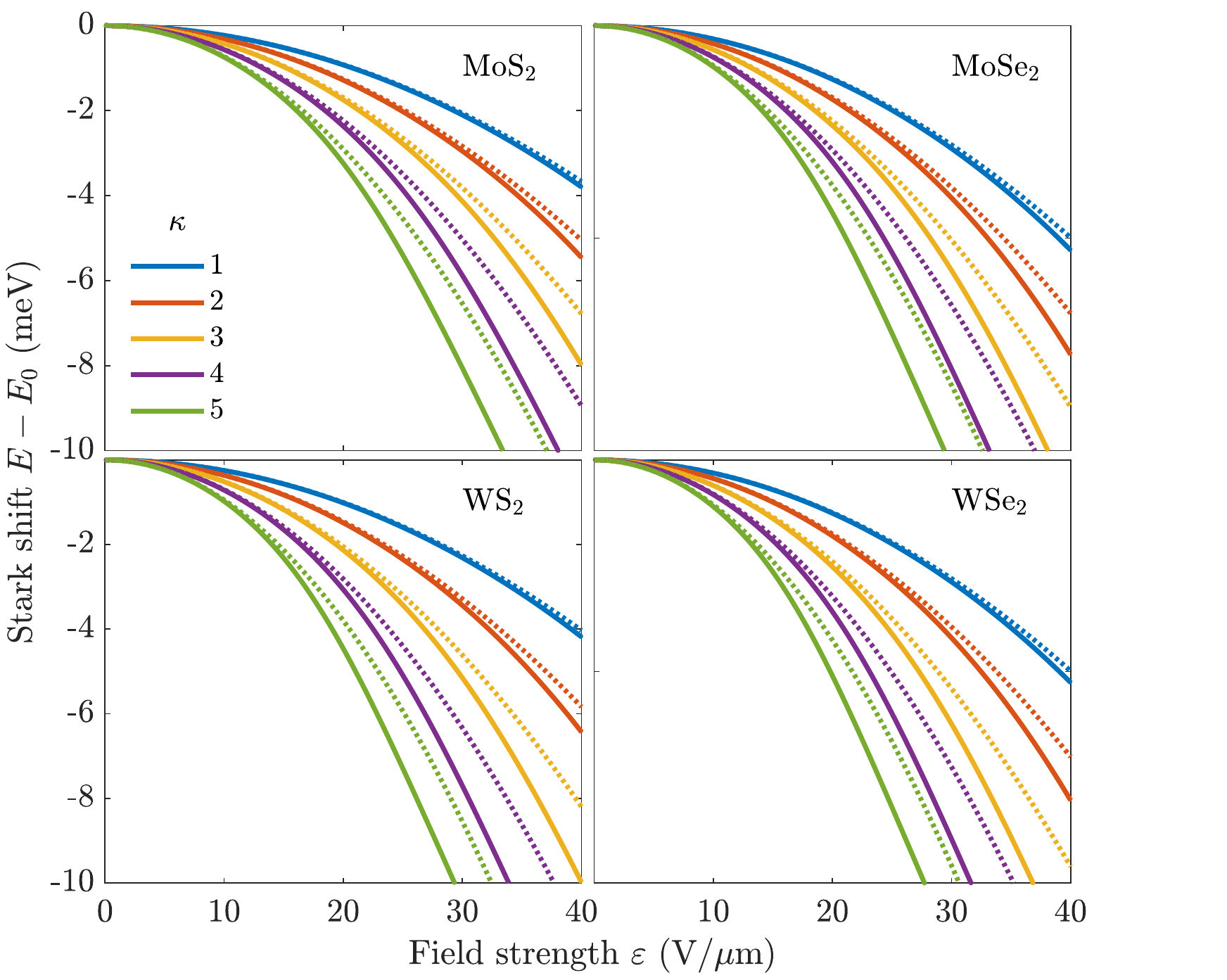}
	\caption{Exciton Stark shift for four important TMDs in various dielectric environments. The solid lines correspond to the real part of the resonance energy, while the dotted lines show to the shift predicted by perturbation theory $E-E_0 \approx -\alpha \varepsilon^2/2$. }\label{fig:Stark_compare}
\end{figure}

\begin{table}[b]
	\caption{Exciton polarizability $\alpha$ for various TMDs in different dielectric environments in units of $10^{-18}\mathrm{eV}\left(\mathrm{m/V}\right)^2$ calculated from \cref{eq:pol} with $\psi_0$ and $\psi_1$ expanded in an FE basis with a spacing of $1$ a.u..}
	\centering
	\begin{tabular}{c ccc c ccc c ccc c ccc}
		\hline\hline\\[-0.2cm]
		\phantom &\multicolumn{1}{ c }{MoS$_2$} 
		&\phantom&\multicolumn{1}{ c }{MoSe$_2$}
		&\phantom&\multicolumn{1}{ c }{WS$_2$}
		&\phantom&\multicolumn{1}{ c }{WSe$_2$}\\
		\cmidrule{2-2}  \cmidrule{4-4} \cmidrule{6-6} \cmidrule{8-8}\\[-0.3cm]
		$\kappa$ &$\alpha$  & \phantom 
		&$\alpha$   &\phantom
		&$\alpha$   &\phantom
		&$\alpha$    
		\\ [0.1cm]\hline\\[-0.2cm]
		$1$ & $4.59$  &&  $6.24$  &&  $5.04$ &&   $6.24$  \\[0.1cm]
		$2$ & $6.31$ &&   $8.46$  &&  $7.30$   && $8.78$  \\[0.1cm]
		$3$ & $8.48$ &&  $11.22$ &&  $10.25$  && $12.02$   \\[0.1cm]
		$4$ & $11.18$  && $14.63$  && $14.09$ &&  $16.14$   \\[0.1cm]
		$5$ & $14.54$ &&  $18.81$  && $19.00$  && $21.32$   \\[0.1cm]
		
	\end{tabular}\label{tbl:excitonpol}
\end{table}
Applying perturbation theory to the ground state of a system with cylindrical symmetry leads to the well known result
\begin{align}
E = E_0 - \frac{1}{2}\alpha \varepsilon^2 + O\left(\varepsilon^4\right)\thinspace,\label{eq:perteng2}
\end{align}
where $\alpha$ is the static polarizability. A shortcoming of perturbation theory is that it predicts the energy as a function of field strength to be purely real, which, as seen in the previous sections, is obviously not correct for a system where dissociation is possible. Nevertheless, the non-perturbative behavior of the resonance energy can be reproduced by utilizing the first few perturbation coefficients together with the hypergeometric resummation technique \cite{Mera2015}. This approach was used in Ref. \cite{Pedersen2016Stark} with great success for low-dimensional hydrogen. In the present section, we wish to analyze to what degree the change in the real part of the resonance energy, i.e. the exciton Stark shift, can be predicted by standard second-order perturbation theory. To this end, we calculate the exciton polarizability given by 
\begin{align}
\alpha = -2\left<\psi_0\right|r\cos\theta\left|\psi_1\right>\thinspace,\label{eq:pol}
\end{align}
where the first order correction $\psi_1$ is a solution to the Dalgarno-Lewis \cite{Dalgarno1955} equation
\begin{align}
\left[-\frac{1}{2\mu}\nabla^2-w\left(\kappa \boldsymbol{r}\right)-E_0\right]\psi_1 = -r\cos\theta \psi_0\thinspace,\label{eq:psi1}
\end{align}
and will therefore be of the form $\psi_1 = \cos\theta f\left(r\right)$, where $f$ is a purely radial function. Expanding $\psi_0$ and $\psi_1$ in a finite element basis (without complex scaling), as described in \cref{app:numerical}, \cref{eq:psi1} can be solved and the polarizability found (for alternative methods of finding the polarizability, see Ref. \cite{Pedersen2016ExcitonStark}). The exciton polarizability for various TMDs in different environments can be found in \cref{tbl:excitonpol}, and \cref{fig:Stark_compare} shows a comparison between the shift in the real part of the complex resonance energy and the perturbation series in \cref{eq:perteng2}. Evidently, a good agreement is found in the weak field region. Furthermore, excitons in environments with large dielectric screening begin to deviate from their second-order expansion for weaker fields than their free-space counterparts. This is to be expected, as the binding energies of excitons with heavily screened interactions are lower and the characteristic fields of these excitons are therefore weaker.

\section{Summary}

In the present work, electric field induced dissociation of TMD excitons has been investigated using both numerical and analytical approaches. The dissociation rates as functions of the in-plane field strength for excitons in monolayer MoS$_2$, MoSe$_2$, WS$_2$, and WSe$_2$ in various screening environments have been obtained. In particular, difficulties associated with dissociation rates in weak electric fields have been addressed and resolved. In this regard, an efficient numerical method capable of computing dissociation rates for a wide range of fields has been introduced. As the field becomes sufficiently weak, any numerical method with finite precision arithmetic breaks down, which calls for a different approach. We demonstrate that an analytical weak-field approximation is valid in this region, which makes the weak-field dissociation rates readily available for arbitrarily weak fields. Finally, the exciton Stark shift has been analyzed and compared to the results of second order perturbation theory. 

\begin{acknowledgments}
The authors gratefully acknowledge financial support by the Center for Nanostructured Graphene (CNG), which is sponsored by the Danish National Research Foundation, Project No. DNRF103. Additionally, T.G.P. is supported by the QUSCOPE Center, sponsored by the Villum Foundation.
\end{acknowledgments}

\appendix
\section{NUMERICAL PROCEDURE}\label{app:numerical}

To implement the finite element (FE) approach, we first divide the radial grid into $N$ segments $\left[r_{n-1},r_n\right]$ for $n=1,...,N$. Following the procedure in Ref. \cite{Scrinzi2010}, we introduce a set of $p_n$ linearly independent functions $h_i^{\left(n\right)}$ where $i=1,...,p_n$ on each segment. These functions are then transformed into a different set of functions $f_i^{\left(n\right)}$, $i=1,...,p_n$, that vanish at the segment boundaries, except for the first and last function, which are required to equal unity at the lower and upper element boundaries, respectively. To summarize,
\begin{align}
&f_i^{\left(n\right)}\left(r_{n-1}\right)=f_i^{\left(n\right)}\left(r_n\right)=0\thinspace,\\
\mathrm{except} \quad &f_1^{\left(n\right)}\left(r_{n-1}\right)=f_{p_n}^{\left(n\right)}\left(r_n\right)=1\thinspace.
\end{align}
We use Legendre polynomials $h_i^{\left(n\right)}\left(r\right)=P_{i-1}\left[y_n\left(r\right)\right]$, where $y_n$ maps $\left[r_{n-1},r_n\right]$ onto $\left[-1,1\right]$, and $h_i^{\left(n\right)}$ is set equal to zero for $r\notin \left[r_{n-1},r_n\right]$. 
Dirichlet boundary conditions are then implemented for some large $r_N\geq R$ by omitting the last function $f_{p_N}^{\left(N\right)}$. The scaling radius $R$ is to be chosen to coincide with an element boundary. Note that if $r_N=R$, no complex scaling is implemented. The eigenstate can now be written as a sum of basis functions
\begin{align}
\psi\left(\boldsymbol{r}\right) = \sum_{m=0}^M\sum_{n=1}^N\sum_{i=1}^{p_n}c_i^{\left(m,n\right)}f_i^{\left(n\right)}\left(r\right)\cos\left(m\theta\right)\thinspace,
\end{align}
where the radial part is resolved using the finite element basis. Due to the cylindrical symmetry of the unperturbed problem the angular dependence of the unperturbed eigenstates are the cylindrical harmonics $e^{im\theta}$\thinspace. The angular part of the eigenstate is therefore resolved efficiently using a basis of cosine functions. To ensure continuity across the segment boundaries, we enforce
\begin{align}
c_{p_{n-1}}^{\left(m,n-1\right)} = c_{1}^{\left(m,n\right)}\thinspace, \quad n=2,...,N\thinspace.\label{eq:cont}
\end{align}
To evaluate the radial part of the matrix elements we use the Legendre quadrature rule \cite{Krylov2005}. 

We proceed by providing a recipe for constructing the overlap and Hamilton matrix, and refer the interested reader to Refs. \cite{Scrinzi2010,McCurdy2004} and references therein for more details on the mathematical background. It is convenient to first construct segmentwise matrices containing only the radial part of the matrix elements. For segments with $r_n\leq R$, the procedure is familiar and, as an example, the radial segmentwise overlap matrices are given by
\begin{align}
S^{\left(n\right)}_{ij} &= \int_{0}^{\infty}f_i^{\left(n\right)}\left(r\right)f_j^{\left(n\right)}\left(r\right)rdr \\
&\approx \sum_{k=1}^K f_i^{\left(n\right)}\left(r_k^{\left(n\right)}\right)f_j^{\left(n\right)}\left(r_k^{\left(n\right)}\right)r_k^{\left(n\right)}w_k^{\left(n\right)}\thinspace,
\end{align}
where $w_k^{\left(n\right)}$ and $r_k^{\left(n\right)}$ are the quadrature weights and sample points for the $n$th segment respectively. For segments with $r_{n-1}\geq R$, the radial coordinate is transformed according to \cref{eq:ECStrans} and the matrix elements must be modified accordingly. The integral element $dr$ must be multiplied by $e^{i\phi}$ and $r$ must be replaced by the transformation in \cref{eq:ECStrans}, all the while keeping the argument of the basis functions unchanged. As an example, the segmentwise overlap matrix becomes
\begin{multline}
S^{\left(n\right)}_{ij} \approx \sum_{k=1}^K\left\{ f_i^{\left(n\right)}\left(r_k^{\left(n\right)}\right)f_j^{\left(n\right)}\left(r_k^{\left(n\right)}\right)\right.\\\left.\times\left[R + \left(r_k^{\left(n\right)}-R\right)e^{i\phi}\right]w_k^{\left(n\right)}e^{i\phi}\right\}\thinspace.
\end{multline}
The segmentwise matrices are then collected into the complete radial overlap matrix $\mathbf{S}_r$ such that the last row and column of each segmentwise matrix overlaps with the first row and column of the next (see Ref. \cite{Scrinzi2010} for a visual demonstration). This conveniently enforces \cref{eq:cont}. The complete overlap matrix $\mathbf{S}$ is then a block diagonal matrix with blocks consisting of $\pi\left(1 + \delta_{m0}\right)\mathbf{S}_r$ for $m=0,...M$. The Hamilton matrix can be constructed in a similar manner, keeping in mind that $\frac{d}{dr}f_i$ should be replaced by $e^{-i\phi}\frac{d}{dr}f_i$ for segments outside the scaling radius. The transformed Wannier equation is then readily solved as a matrix eigenvalue problem.
\\

\section{COMPUTATIONAL PROCEDURE FOR THE ASYMPTOTIC COEFFICIENT}\label{app:numerov}

Grid-based finite-difference methods (FDMs) are able to efficiently reproduce the correct behavior of the wave function for all values of $r$, as long as a dense enough grid is used. Numerov's method is a fourth-order FDM, on par with the fourth-order Runge-Kutta method. However, the advantage is that it is simpler to implement. The ground-state wave function can be presented on the form $\psi_0\left(r\right) = r^{-1/2}P\left(r\right)$, which transforms \cref{eq:SimplifiedWannier} (with $\varepsilon=0$) to the differential equation
\begin{align}
	\frac{d^2P\left(r\right)}{dr^2} + g\left(r\right)P\left(r\right) = 0\thinspace,
\end{align}
where
\begin{align}
	g\left(r\right) = \frac{1}{4r^2}+2\left[E_0+w\left(r\right)\right]\thinspace.
\end{align}
We now assume $E_0$ is known (it can easily be calculate by e.g. diagonalizing a Gaussian basis or performing a variational calculation). Numerov's method then reduces this equation to the finite difference equation
\begin{align}
	f_{n-1}P_{n-1} = \left(12 -10f_n\right)P_n -f_{n+1}P_{n+1}\thinspace,
\end{align}
where $P_n = P(r_n)$ and $f_n = 1 + \left(\Delta r\right)^2g_n/12$ with $g_n = g\left(r_n\right)$. The $N+1$ discrete points $r_n$ are defined as $r_n = n\Delta r$, where $n=0,...,N$ and $\Delta r = r_N/N$. The two initial points are then chosen as $P_{N} =0$ for some large $r_N$ and to comply with \cref{eq:asymp} for $r_{N-1}$. Integrating towards $r=0$ then yields $P$ at all $r_n$. The fitting procedure described in the main text is then implemented by fitting \cref{eq:C0as} to \cref{eq:fitas} in a region $r\in\left[\left(j-1\right)40,j40\right]$, where $j=1,2...$, until convergence to $4$ significant digits. The same $P$ is then used in \cref{eq:Cint} and the agreeing significant digits (up to fourth order) are presented in \cref{tbl:1}.

\bibliography{litt}

\begin{thebibliography}{50}%
\makeatletter
\providecommand \@ifxundefined [1]{%
 \@ifx{#1\undefined}
}%
\providecommand \@ifnum [1]{%
 \ifnum #1\expandafter \@firstoftwo
 \else \expandafter \@secondoftwo
 \fi
}%
\providecommand \@ifx [1]{%
 \ifx #1\expandafter \@firstoftwo
 \else \expandafter \@secondoftwo
 \fi
}%
\providecommand \natexlab [1]{#1}%
\providecommand \enquote  [1]{``#1''}%
\providecommand \bibnamefont  [1]{#1}%
\providecommand \bibfnamefont [1]{#1}%
\providecommand \citenamefont [1]{#1}%
\providecommand \href@noop [0]{\@secondoftwo}%
\providecommand \href [0]{\begingroup \@sanitize@url \@href}%
\providecommand \@href[1]{\@@startlink{#1}\@@href}%
\providecommand \@@href[1]{\endgroup#1\@@endlink}%
\providecommand \@sanitize@url [0]{\catcode `\\12\catcode `\$12\catcode
  `\&12\catcode `\#12\catcode `\^12\catcode `\_12\catcode `\%12\relax}%
\providecommand \@@startlink[1]{}%
\providecommand \@@endlink[0]{}%
\providecommand \url  [0]{\begingroup\@sanitize@url \@url }%
\providecommand \@url [1]{\endgroup\@href {#1}{\urlprefix }}%
\providecommand \urlprefix  [0]{URL }%
\providecommand \Eprint [0]{\href }%
\providecommand \doibase [0]{http://dx.doi.org/}%
\providecommand \selectlanguage [0]{\@gobble}%
\providecommand \bibinfo  [0]{\@secondoftwo}%
\providecommand \bibfield  [0]{\@secondoftwo}%
\providecommand \translation [1]{[#1]}%
\providecommand \BibitemOpen [0]{}%
\providecommand \bibitemStop [0]{}%
\providecommand \bibitemNoStop [0]{.\EOS\space}%
\providecommand \EOS [0]{\spacefactor3000\relax}%
\providecommand \BibitemShut  [1]{\csname bibitem#1\endcsname}%
\let\auto@bib@innerbib\@empty
\bibitem [{\citenamefont {Wang}\ \emph {et~al.}(2015)\citenamefont {Wang},
  \citenamefont {Zhang}, \citenamefont {Chan}, \citenamefont {Tiwari},\ and\
  \citenamefont {Rana}}]{Wang2015}%
  \BibitemOpen
  \bibfield  {author} {\bibinfo {author} {\bibfnamefont {H.}~\bibnamefont
  {Wang}}, \bibinfo {author} {\bibfnamefont {C.}~\bibnamefont {Zhang}},
  \bibinfo {author} {\bibfnamefont {W.}~\bibnamefont {Chan}}, \bibinfo {author}
  {\bibfnamefont {S.}~\bibnamefont {Tiwari}}, \ and\ \bibinfo {author}
  {\bibfnamefont {F.}~\bibnamefont {Rana}},\ }\href {\doibase
  10.1038/ncomms9831} {\bibfield  {journal} {\bibinfo  {journal} {Nat.
  Commun.}\ }\textbf {\bibinfo {volume} {6}},\ \bibinfo {pages} {8831}
  (\bibinfo {year} {2015})}\BibitemShut {NoStop}%
\bibitem [{\citenamefont {Lopez-Sanchez}\ \emph {et~al.}(2013)\citenamefont
  {Lopez-Sanchez}, \citenamefont {Lembke}, \citenamefont {Kayci}, \citenamefont
  {Radenovic},\ and\ \citenamefont {Kis}}]{Lopez2013}%
  \BibitemOpen
  \bibfield  {author} {\bibinfo {author} {\bibfnamefont {O.}~\bibnamefont
  {Lopez-Sanchez}}, \bibinfo {author} {\bibfnamefont {D.}~\bibnamefont
  {Lembke}}, \bibinfo {author} {\bibfnamefont {M.}~\bibnamefont {Kayci}},
  \bibinfo {author} {\bibfnamefont {A.}~\bibnamefont {Radenovic}}, \ and\
  \bibinfo {author} {\bibfnamefont {A.}~\bibnamefont {Kis}},\ }\href {\doibase
  10.1038/nnano.2013.100} {\bibfield  {journal} {\bibinfo  {journal} {Nat.
  Nanotechnol.}\ }\textbf {\bibinfo {volume} {8}},\ \bibinfo {pages} {449}
  (\bibinfo {year} {2013})}\BibitemShut {NoStop}%
\bibitem [{\citenamefont {Yin}\ \emph {et~al.}(2012)\citenamefont {Yin},
  \citenamefont {Li}, \citenamefont {Li}, \citenamefont {Jiang}, \citenamefont
  {Shi}, \citenamefont {Sun}, \citenamefont {Lu}, \citenamefont {Zhang},
  \citenamefont {Chen},\ and\ \citenamefont {Zhang}}]{Yin2012}%
  \BibitemOpen
  \bibfield  {author} {\bibinfo {author} {\bibfnamefont {Z.}~\bibnamefont
  {Yin}}, \bibinfo {author} {\bibfnamefont {H.}~\bibnamefont {Li}}, \bibinfo
  {author} {\bibfnamefont {H.}~\bibnamefont {Li}}, \bibinfo {author}
  {\bibfnamefont {L.}~\bibnamefont {Jiang}}, \bibinfo {author} {\bibfnamefont
  {Y.}~\bibnamefont {Shi}}, \bibinfo {author} {\bibfnamefont {Y.}~\bibnamefont
  {Sun}}, \bibinfo {author} {\bibfnamefont {G.}~\bibnamefont {Lu}}, \bibinfo
  {author} {\bibfnamefont {Q.}~\bibnamefont {Zhang}}, \bibinfo {author}
  {\bibfnamefont {X.}~\bibnamefont {Chen}}, \ and\ \bibinfo {author}
  {\bibfnamefont {H.}~\bibnamefont {Zhang}},\ }\href {\doibase
  10.1021/nn2024557} {\bibfield  {journal} {\bibinfo  {journal} {ACS Nano}\
  }\textbf {\bibinfo {volume} {6}},\ \bibinfo {pages} {74} (\bibinfo {year}
  {2012})}\BibitemShut {NoStop}%
\bibitem [{\citenamefont {Withers}\ \emph {et~al.}(2015)\citenamefont
  {Withers}, \citenamefont {Pozo-Zamudio}, \citenamefont {Mishchenko},
  \citenamefont {Rooney}, \citenamefont {Gholinia}, \citenamefont {Watanabe},
  \citenamefont {Taniguchi}, \citenamefont {Haigh}, \citenamefont {Geim},
  \citenamefont {Tartakovskii},\ and\ \citenamefont {Novoselov}}]{Withers2015}%
  \BibitemOpen
  \bibfield  {author} {\bibinfo {author} {\bibfnamefont {F.}~\bibnamefont
  {Withers}}, \bibinfo {author} {\bibfnamefont {O.~D.}\ \bibnamefont
  {Pozo-Zamudio}}, \bibinfo {author} {\bibfnamefont {A.}~\bibnamefont
  {Mishchenko}}, \bibinfo {author} {\bibfnamefont {A.~P.}\ \bibnamefont
  {Rooney}}, \bibinfo {author} {\bibfnamefont {A.}~\bibnamefont {Gholinia}},
  \bibinfo {author} {\bibfnamefont {K.}~\bibnamefont {Watanabe}}, \bibinfo
  {author} {\bibfnamefont {T.}~\bibnamefont {Taniguchi}}, \bibinfo {author}
  {\bibfnamefont {S.~J.}\ \bibnamefont {Haigh}}, \bibinfo {author}
  {\bibfnamefont {A.~K.}\ \bibnamefont {Geim}}, \bibinfo {author}
  {\bibfnamefont {A.~I.}\ \bibnamefont {Tartakovskii}}, \ and\ \bibinfo
  {author} {\bibfnamefont {K.~S.}\ \bibnamefont {Novoselov}},\ }\href {\doibase
  10.1038/nmat4205} {\bibfield  {journal} {\bibinfo  {journal} {Nat. Mater.}\
  }\textbf {\bibinfo {volume} {14}},\ \bibinfo {pages} {301} (\bibinfo {year}
  {2015})}\BibitemShut {NoStop}%
\bibitem [{\citenamefont {Lopez-Sanchez}\ \emph {et~al.}(2014)\citenamefont
  {Lopez-Sanchez}, \citenamefont {Alarcon~Llado}, \citenamefont {Koman},
  \citenamefont {Fontcuberta~i Morral}, \citenamefont {Radenovic},\ and\
  \citenamefont {Kis}}]{Lopez2014}%
  \BibitemOpen
  \bibfield  {author} {\bibinfo {author} {\bibfnamefont {O.}~\bibnamefont
  {Lopez-Sanchez}}, \bibinfo {author} {\bibfnamefont {E.}~\bibnamefont
  {Alarcon~Llado}}, \bibinfo {author} {\bibfnamefont {V.}~\bibnamefont
  {Koman}}, \bibinfo {author} {\bibfnamefont {A.}~\bibnamefont {Fontcuberta~i
  Morral}}, \bibinfo {author} {\bibfnamefont {A.}~\bibnamefont {Radenovic}}, \
  and\ \bibinfo {author} {\bibfnamefont {A.}~\bibnamefont {Kis}},\ }\href
  {\doibase 10.1021/nn500480u} {\bibfield  {journal} {\bibinfo  {journal} {ACS
  Nano}\ }\textbf {\bibinfo {volume} {8}},\ \bibinfo {pages} {3042} (\bibinfo
  {year} {2014})}\BibitemShut {NoStop}%
\bibitem [{\citenamefont {Bernardi}\ \emph {et~al.}(2013)\citenamefont
  {Bernardi}, \citenamefont {Palummo},\ and\ \citenamefont
  {Grossman}}]{Bernardi2013}%
  \BibitemOpen
  \bibfield  {author} {\bibinfo {author} {\bibfnamefont {M.}~\bibnamefont
  {Bernardi}}, \bibinfo {author} {\bibfnamefont {M.}~\bibnamefont {Palummo}}, \
  and\ \bibinfo {author} {\bibfnamefont {J.~C.}\ \bibnamefont {Grossman}},\
  }\href {\doibase 10.1021/nl401544y} {\bibfield  {journal} {\bibinfo
  {journal} {Nano Lett.}\ }\textbf {\bibinfo {volume} {13}},\ \bibinfo {pages}
  {3664} (\bibinfo {year} {2013})}\BibitemShut {NoStop}%
\bibitem [{\citenamefont {Du}\ \emph {et~al.}(2010)\citenamefont {Du},
  \citenamefont {Guo}, \citenamefont {Wang}, \citenamefont {Zeng},
  \citenamefont {Chen},\ and\ \citenamefont {Liu}}]{Guodong2010}%
  \BibitemOpen
  \bibfield  {author} {\bibinfo {author} {\bibfnamefont {G.}~\bibnamefont
  {Du}}, \bibinfo {author} {\bibfnamefont {Z.}~\bibnamefont {Guo}}, \bibinfo
  {author} {\bibfnamefont {S.}~\bibnamefont {Wang}}, \bibinfo {author}
  {\bibfnamefont {R.}~\bibnamefont {Zeng}}, \bibinfo {author} {\bibfnamefont
  {Z.}~\bibnamefont {Chen}}, \ and\ \bibinfo {author} {\bibfnamefont
  {H.}~\bibnamefont {Liu}},\ }\href {\doibase 10.1039/B920277C} {\bibfield
  {journal} {\bibinfo  {journal} {Chem. Commun.}\ }\textbf {\bibinfo {volume}
  {46}},\ \bibinfo {pages} {1106} (\bibinfo {year} {2010})}\BibitemShut
  {NoStop}%
\bibitem [{\citenamefont {Chhowalla}\ \emph {et~al.}(2013)\citenamefont
  {Chhowalla}, \citenamefont {Shin}, \citenamefont {Eda}, \citenamefont {Li},
  \citenamefont {Loh},\ and\ \citenamefont {Zhang}}]{Chhowalla2013}%
  \BibitemOpen
  \bibfield  {author} {\bibinfo {author} {\bibfnamefont {M.}~\bibnamefont
  {Chhowalla}}, \bibinfo {author} {\bibfnamefont {H.~S.}\ \bibnamefont {Shin}},
  \bibinfo {author} {\bibfnamefont {G.}~\bibnamefont {Eda}}, \bibinfo {author}
  {\bibfnamefont {L.-J.}\ \bibnamefont {Li}}, \bibinfo {author} {\bibfnamefont
  {K.~P.}\ \bibnamefont {Loh}}, \ and\ \bibinfo {author} {\bibfnamefont
  {H.}~\bibnamefont {Zhang}},\ }\href {\doibase 10.1038/nchem.1589} {\bibfield
  {journal} {\bibinfo  {journal} {Nat. Chem.}\ }\textbf {\bibinfo {volume}
  {5}},\ \bibinfo {pages} {263} (\bibinfo {year} {2013})}\BibitemShut {NoStop}%
\bibitem [{\citenamefont {Soon}\ and\ \citenamefont {Loh}(2007)}]{Soon2007}%
  \BibitemOpen
  \bibfield  {author} {\bibinfo {author} {\bibfnamefont {J.~M.}\ \bibnamefont
  {Soon}}\ and\ \bibinfo {author} {\bibfnamefont {K.~P.}\ \bibnamefont {Loh}},\
  }\href {\doibase 10.1149/1.2778851} {\bibfield  {journal} {\bibinfo
  {journal} {Electrochem. Solid State Lett.}\ }\textbf {\bibinfo {volume}
  {10}},\ \bibinfo {pages} {A250} (\bibinfo {year} {2007})}\BibitemShut
  {NoStop}%
\bibitem [{\citenamefont {Ramasubramaniam}(2012)}]{Ramasubramaniam2012}%
  \BibitemOpen
  \bibfield  {author} {\bibinfo {author} {\bibfnamefont {A.}~\bibnamefont
  {Ramasubramaniam}},\ }\href {\doibase 10.1103/PhysRevB.86.115409} {\bibfield
  {journal} {\bibinfo  {journal} {Phys. Rev. B}\ }\textbf {\bibinfo {volume}
  {86}},\ \bibinfo {pages} {115409} (\bibinfo {year} {2012})}\BibitemShut
  {NoStop}%
\bibitem [{\citenamefont {Latini}\ \emph {et~al.}(2015)\citenamefont {Latini},
  \citenamefont {Olsen},\ and\ \citenamefont {Thygesen}}]{Latini2015}%
  \BibitemOpen
  \bibfield  {author} {\bibinfo {author} {\bibfnamefont {S.}~\bibnamefont
  {Latini}}, \bibinfo {author} {\bibfnamefont {T.}~\bibnamefont {Olsen}}, \
  and\ \bibinfo {author} {\bibfnamefont {K.~S.}\ \bibnamefont {Thygesen}},\
  }\href {\doibase 10.1103/PhysRevB.92.245123} {\bibfield  {journal} {\bibinfo
  {journal} {Phys. Rev. B}\ }\textbf {\bibinfo {volume} {92}},\ \bibinfo
  {pages} {245123} (\bibinfo {year} {2015})}\BibitemShut {NoStop}%
\bibitem [{\citenamefont {Berkelbach}\ \emph {et~al.}(2013)\citenamefont
  {Berkelbach}, \citenamefont {Hybertsen},\ and\ \citenamefont
  {Reichman}}]{Berkelbach2013}%
  \BibitemOpen
  \bibfield  {author} {\bibinfo {author} {\bibfnamefont {T.~C.}\ \bibnamefont
  {Berkelbach}}, \bibinfo {author} {\bibfnamefont {M.~S.}\ \bibnamefont
  {Hybertsen}}, \ and\ \bibinfo {author} {\bibfnamefont {D.~R.}\ \bibnamefont
  {Reichman}},\ }\href {\doibase 10.1103/PhysRevB.88.045318} {\bibfield
  {journal} {\bibinfo  {journal} {Phys. Rev. B}\ }\textbf {\bibinfo {volume}
  {88}},\ \bibinfo {pages} {045318} (\bibinfo {year} {2013})}\BibitemShut
  {NoStop}%
\bibitem [{\citenamefont {Pedersen}\ \emph
  {et~al.}(2016{\natexlab{a}})\citenamefont {Pedersen}, \citenamefont {Latini},
  \citenamefont {Thygesen}, \citenamefont {Mera},\ and\ \citenamefont
  {Nikolić}}]{Pedersen2016ExcitonIonization}%
  \BibitemOpen
  \bibfield  {author} {\bibinfo {author} {\bibfnamefont {T.~G.}\ \bibnamefont
  {Pedersen}}, \bibinfo {author} {\bibfnamefont {S.}~\bibnamefont {Latini}},
  \bibinfo {author} {\bibfnamefont {K.~S.}\ \bibnamefont {Thygesen}}, \bibinfo
  {author} {\bibfnamefont {H.}~\bibnamefont {Mera}}, \ and\ \bibinfo {author}
  {\bibfnamefont {B.~K.}\ \bibnamefont {Nikolić}},\ }\href
  {http://stacks.iop.org/1367-2630/18/i=7/a=073043} {\bibfield  {journal}
  {\bibinfo  {journal} {New J. Phys.}\ }\textbf {\bibinfo {volume} {18}},\
  \bibinfo {pages} {073043} (\bibinfo {year} {2016}{\natexlab{a}})}\BibitemShut
  {NoStop}%
\bibitem [{\citenamefont {Haastrup}\ \emph {et~al.}(2016)\citenamefont
  {Haastrup}, \citenamefont {Latini}, \citenamefont {Bolotin},\ and\
  \citenamefont {Thygesen}}]{Haastrup2016}%
  \BibitemOpen
  \bibfield  {author} {\bibinfo {author} {\bibfnamefont {S.}~\bibnamefont
  {Haastrup}}, \bibinfo {author} {\bibfnamefont {S.}~\bibnamefont {Latini}},
  \bibinfo {author} {\bibfnamefont {K.}~\bibnamefont {Bolotin}}, \ and\
  \bibinfo {author} {\bibfnamefont {K.~S.}\ \bibnamefont {Thygesen}},\ }\href
  {\doibase 10.1103/PhysRevB.94.041401} {\bibfield  {journal} {\bibinfo
  {journal} {Phys. Rev. B}\ }\textbf {\bibinfo {volume} {94}},\ \bibinfo
  {pages} {041401} (\bibinfo {year} {2016})}\BibitemShut {NoStop}%
\bibitem [{\citenamefont {Massicotte}\ \emph {et~al.}(2018)\citenamefont
  {Massicotte}, \citenamefont {Vialla}, \citenamefont {Schmidt}, \citenamefont
  {B.~Lundeberg}, \citenamefont {Latini}, \citenamefont {Haastrup},
  \citenamefont {Danovich}, \citenamefont {Davydovskaya}, \citenamefont
  {Watanabe}, \citenamefont {Taniguchi}, \citenamefont {I.~Falko},
  \citenamefont {Thygesen}, \citenamefont {G.~Pedersen},\ and\ \citenamefont
  {H.~L.~Koppens}}]{Massicotte2018}%
  \BibitemOpen
  \bibfield  {author} {\bibinfo {author} {\bibfnamefont {M.}~\bibnamefont
  {Massicotte}}, \bibinfo {author} {\bibfnamefont {F.}~\bibnamefont {Vialla}},
  \bibinfo {author} {\bibfnamefont {P.}~\bibnamefont {Schmidt}}, \bibinfo
  {author} {\bibfnamefont {M.}~\bibnamefont {B.~Lundeberg}}, \bibinfo {author}
  {\bibfnamefont {S.}~\bibnamefont {Latini}}, \bibinfo {author} {\bibfnamefont
  {S.}~\bibnamefont {Haastrup}}, \bibinfo {author} {\bibfnamefont
  {M.}~\bibnamefont {Danovich}}, \bibinfo {author} {\bibfnamefont
  {D.}~\bibnamefont {Davydovskaya}}, \bibinfo {author} {\bibfnamefont
  {K.}~\bibnamefont {Watanabe}}, \bibinfo {author} {\bibfnamefont
  {T.}~\bibnamefont {Taniguchi}}, \bibinfo {author} {\bibfnamefont
  {V.}~\bibnamefont {I.~Falko}}, \bibinfo {author} {\bibfnamefont
  {K.}~\bibnamefont {Thygesen}}, \bibinfo {author} {\bibfnamefont
  {T.}~\bibnamefont {G.~Pedersen}}, \ and\ \bibinfo {author} {\bibfnamefont
  {F.}~\bibnamefont {H.~L.~Koppens}},\ }\href {\doibase
  10.1038/s41467-018-03864-y} {\bibfield  {journal} {\bibinfo  {journal} {Nat.
  Commun.}\ }\textbf {\bibinfo {volume} {9}},\ \bibinfo {pages} {1633}
  (\bibinfo {year} {2018})}\BibitemShut {NoStop}%
\bibitem [{\citenamefont {Wannier}(1937)}]{Wannier1937}%
  \BibitemOpen
  \bibfield  {author} {\bibinfo {author} {\bibfnamefont {G.~H.}\ \bibnamefont
  {Wannier}},\ }\href {\doibase 10.1103/PhysRev.52.191} {\bibfield  {journal}
  {\bibinfo  {journal} {Phys. Rev.}\ }\textbf {\bibinfo {volume} {52}},\
  \bibinfo {pages} {191} (\bibinfo {year} {1937})}\BibitemShut {NoStop}%
\bibitem [{\citenamefont {Lederman}\ and\ \citenamefont
  {Dow}(1976)}]{Lederman1976}%
  \BibitemOpen
  \bibfield  {author} {\bibinfo {author} {\bibfnamefont {F.~L.}\ \bibnamefont
  {Lederman}}\ and\ \bibinfo {author} {\bibfnamefont {J.~D.}\ \bibnamefont
  {Dow}},\ }\href {\doibase 10.1103/PhysRevB.13.1633} {\bibfield  {journal}
  {\bibinfo  {journal} {Phys. Rev. B}\ }\textbf {\bibinfo {volume} {13}},\
  \bibinfo {pages} {1633} (\bibinfo {year} {1976})}\BibitemShut {NoStop}%
\bibitem [{\citenamefont {Pedersen}\ \emph
  {et~al.}(2016{\natexlab{b}})\citenamefont {Pedersen}, \citenamefont {Mera},\
  and\ \citenamefont {Nikoli\ifmmode~\acute{c}\else
  \'{c}\fi{}}}]{Pedersen2016Stark}%
  \BibitemOpen
  \bibfield  {author} {\bibinfo {author} {\bibfnamefont {T.~G.}\ \bibnamefont
  {Pedersen}}, \bibinfo {author} {\bibfnamefont {H.}~\bibnamefont {Mera}}, \
  and\ \bibinfo {author} {\bibfnamefont {B.~K.}\ \bibnamefont
  {Nikoli\ifmmode~\acute{c}\else \'{c}\fi{}}},\ }\href {\doibase
  10.1103/PhysRevA.93.013409} {\bibfield  {journal} {\bibinfo  {journal} {Phys.
  Rev. A}\ }\textbf {\bibinfo {volume} {93}},\ \bibinfo {pages} {013409}
  (\bibinfo {year} {2016}{\natexlab{b}})}\BibitemShut {NoStop}%
\bibitem [{\citenamefont {Balslev}\ and\ \citenamefont
  {Combes}(1971)}]{Balslev1971}%
  \BibitemOpen
  \bibfield  {author} {\bibinfo {author} {\bibfnamefont {E.}~\bibnamefont
  {Balslev}}\ and\ \bibinfo {author} {\bibfnamefont {J.~M.}\ \bibnamefont
  {Combes}},\ }\href {https://projecteuclid.org:443/euclid.cmp/1103857513}
  {\bibfield  {journal} {\bibinfo  {journal} {Comm. Math. Phys.}\ }\textbf
  {\bibinfo {volume} {22}},\ \bibinfo {pages} {280} (\bibinfo {year}
  {1971})}\BibitemShut {NoStop}%
\bibitem [{\citenamefont {Aguilar}\ and\ \citenamefont
  {Combes}(1971)}]{Aguilar1971}%
  \BibitemOpen
  \bibfield  {author} {\bibinfo {author} {\bibfnamefont {J.}~\bibnamefont
  {Aguilar}}\ and\ \bibinfo {author} {\bibfnamefont {J.~M.}\ \bibnamefont
  {Combes}},\ }\href {https://projecteuclid.org:443/euclid.cmp/1103857512}
  {\bibfield  {journal} {\bibinfo  {journal} {Comm. Math. Phys.}\ }\textbf
  {\bibinfo {volume} {22}},\ \bibinfo {pages} {269} (\bibinfo {year}
  {1971})}\BibitemShut {NoStop}%
\bibitem [{\citenamefont {Tolstikhin}\ \emph {et~al.}(2011)\citenamefont
  {Tolstikhin}, \citenamefont {Morishita},\ and\ \citenamefont
  {Madsen}}]{Tolstikhin2011}%
  \BibitemOpen
  \bibfield  {author} {\bibinfo {author} {\bibfnamefont {O.~I.}\ \bibnamefont
  {Tolstikhin}}, \bibinfo {author} {\bibfnamefont {T.}~\bibnamefont
  {Morishita}}, \ and\ \bibinfo {author} {\bibfnamefont {L.~B.}\ \bibnamefont
  {Madsen}},\ }\href {\doibase 10.1103/PhysRevA.84.053423} {\bibfield
  {journal} {\bibinfo  {journal} {Phys. Rev. A}\ }\textbf {\bibinfo {volume}
  {84}},\ \bibinfo {pages} {053423} (\bibinfo {year} {2011})}\BibitemShut
  {NoStop}%
\bibitem [{\citenamefont {Cudazzo}\ \emph {et~al.}(2010)\citenamefont
  {Cudazzo}, \citenamefont {Attaccalite}, \citenamefont {Tokatly},\ and\
  \citenamefont {Rubio}}]{Cudazzo2010}%
  \BibitemOpen
  \bibfield  {author} {\bibinfo {author} {\bibfnamefont {P.}~\bibnamefont
  {Cudazzo}}, \bibinfo {author} {\bibfnamefont {C.}~\bibnamefont
  {Attaccalite}}, \bibinfo {author} {\bibfnamefont {I.~V.}\ \bibnamefont
  {Tokatly}}, \ and\ \bibinfo {author} {\bibfnamefont {A.}~\bibnamefont
  {Rubio}},\ }\href {\doibase 10.1103/PhysRevLett.104.226804} {\bibfield
  {journal} {\bibinfo  {journal} {Phys. Rev. Lett.}\ }\textbf {\bibinfo
  {volume} {104}},\ \bibinfo {pages} {226804} (\bibinfo {year}
  {2010})}\BibitemShut {NoStop}%
\bibitem [{\citenamefont {Keldysh}(1979)}]{Keldysh1979}%
  \BibitemOpen
  \bibfield  {author} {\bibinfo {author} {\bibfnamefont {L.~V.}\ \bibnamefont
  {Keldysh}},\ }\href@noop {} {\bibfield  {journal} {\bibinfo  {journal} {JETP
  Lett.}\ }\textbf {\bibinfo {volume} {29}},\ \bibinfo {pages} {658} (\bibinfo
  {year} {1979})}\BibitemShut {NoStop}%
\bibitem [{\citenamefont {Trolle}\ \emph {et~al.}(2017)\citenamefont {Trolle},
  \citenamefont {Pedersen},\ and\ \citenamefont {Véniard}}]{Trolle2017}%
  \BibitemOpen
  \bibfield  {author} {\bibinfo {author} {\bibfnamefont {M.~L.}\ \bibnamefont
  {Trolle}}, \bibinfo {author} {\bibfnamefont {T.~G.}\ \bibnamefont
  {Pedersen}}, \ and\ \bibinfo {author} {\bibfnamefont {V.}~\bibnamefont
  {Véniard}},\ }\href {\doibase 10.1038/srep39844} {\bibfield  {journal}
  {\bibinfo  {journal} {Sci. Rep.}\ }\textbf {\bibinfo {volume} {7}},\ \bibinfo
  {pages} {39844} (\bibinfo {year} {2017})}\BibitemShut {NoStop}%
\bibitem [{\citenamefont {Abramowitz}(1974)}]{Abramowitz1974}%
  \BibitemOpen
  \bibfield  {author} {\bibinfo {author} {\bibfnamefont {M.}~\bibnamefont
  {Abramowitz}},\ }\href@noop {} {\emph {\bibinfo {title} {Handbook of
  Mathematical Functions, With Formulas, Graphs, and Mathematical Tables}}}\
  (\bibinfo  {publisher} {Dover Publications, Incorporated},\ \bibinfo {year}
  {1974})\BibitemShut {NoStop}%
\bibitem [{\citenamefont {Koch}\ \emph {et~al.}(2006)\citenamefont {Koch},
  \citenamefont {Kira}, \citenamefont {Khitrova},\ and\ \citenamefont
  {M~Gibbs}}]{Koch2006}%
  \BibitemOpen
  \bibfield  {author} {\bibinfo {author} {\bibfnamefont {S.}~\bibnamefont
  {Koch}}, \bibinfo {author} {\bibfnamefont {M.}~\bibnamefont {Kira}}, \bibinfo
  {author} {\bibfnamefont {G.}~\bibnamefont {Khitrova}}, \ and\ \bibinfo
  {author} {\bibfnamefont {H.}~\bibnamefont {M~Gibbs}},\ }\href {\doibase
  10.1038/nmat1658} {\bibfield  {journal} {\bibinfo  {journal} {Nat. Mater.}\
  }\textbf {\bibinfo {volume} {5}},\ \bibinfo {pages} {523} (\bibinfo {year}
  {2006})}\BibitemShut {NoStop}%
\bibitem [{\citenamefont {Palummo}\ \emph {et~al.}(2015)\citenamefont
  {Palummo}, \citenamefont {Bernardi},\ and\ \citenamefont
  {Grossman}}]{Palummo2015}%
  \BibitemOpen
  \bibfield  {author} {\bibinfo {author} {\bibfnamefont {M.}~\bibnamefont
  {Palummo}}, \bibinfo {author} {\bibfnamefont {M.}~\bibnamefont {Bernardi}}, \
  and\ \bibinfo {author} {\bibfnamefont {J.~C.}\ \bibnamefont {Grossman}},\
  }\href {\doibase 10.1021/nl503799t} {\bibfield  {journal} {\bibinfo
  {journal} {Nano Lett.}\ }\textbf {\bibinfo {volume} {15}},\ \bibinfo {pages}
  {2794} (\bibinfo {year} {2015})}\BibitemShut {NoStop}%
\bibitem [{\citenamefont {Wang}\ \emph {et~al.}(2016)\citenamefont {Wang},
  \citenamefont {Zhang}, \citenamefont {Chan}, \citenamefont {Manolatou},
  \citenamefont {Tiwari},\ and\ \citenamefont {Rana}}]{Wang2016}%
  \BibitemOpen
  \bibfield  {author} {\bibinfo {author} {\bibfnamefont {H.}~\bibnamefont
  {Wang}}, \bibinfo {author} {\bibfnamefont {C.}~\bibnamefont {Zhang}},
  \bibinfo {author} {\bibfnamefont {W.}~\bibnamefont {Chan}}, \bibinfo {author}
  {\bibfnamefont {C.}~\bibnamefont {Manolatou}}, \bibinfo {author}
  {\bibfnamefont {S.}~\bibnamefont {Tiwari}}, \ and\ \bibinfo {author}
  {\bibfnamefont {F.}~\bibnamefont {Rana}},\ }\href {\doibase
  10.1103/PhysRevB.93.045407} {\bibfield  {journal} {\bibinfo  {journal} {Phys.
  Rev. B}\ }\textbf {\bibinfo {volume} {93}},\ \bibinfo {pages} {045407}
  (\bibinfo {year} {2016})}\BibitemShut {NoStop}%
\bibitem [{\citenamefont {Poellmann}\ \emph {et~al.}(2015)\citenamefont
  {Poellmann}, \citenamefont {Steinleitner}, \citenamefont {Leierseder},
  \citenamefont {Nagler}, \citenamefont {Plechinger}, \citenamefont {Porer},
  \citenamefont {Bratschitsch}, \citenamefont {Schüller}, \citenamefont
  {Korn},\ and\ \citenamefont {Huber}}]{Poellmann2015}%
  \BibitemOpen
  \bibfield  {author} {\bibinfo {author} {\bibfnamefont {C.}~\bibnamefont
  {Poellmann}}, \bibinfo {author} {\bibfnamefont {P.}~\bibnamefont
  {Steinleitner}}, \bibinfo {author} {\bibfnamefont {U.}~\bibnamefont
  {Leierseder}}, \bibinfo {author} {\bibfnamefont {P.}~\bibnamefont {Nagler}},
  \bibinfo {author} {\bibfnamefont {G.}~\bibnamefont {Plechinger}}, \bibinfo
  {author} {\bibfnamefont {M.}~\bibnamefont {Porer}}, \bibinfo {author}
  {\bibfnamefont {R.}~\bibnamefont {Bratschitsch}}, \bibinfo {author}
  {\bibfnamefont {C.}~\bibnamefont {Schüller}}, \bibinfo {author}
  {\bibfnamefont {T.}~\bibnamefont {Korn}}, \ and\ \bibinfo {author}
  {\bibfnamefont {R.}~\bibnamefont {Huber}},\ }\href {\doibase
  10.1038/nmat4356} {\bibfield  {journal} {\bibinfo  {journal} {Nat. Mater.}\
  }\textbf {\bibinfo {volume} {14}},\ \bibinfo {pages} {889} (\bibinfo {year}
  {2015})}\BibitemShut {NoStop}%
\bibitem [{\citenamefont {Shi}\ \emph {et~al.}(2013)\citenamefont {Shi},
  \citenamefont {Yan}, \citenamefont {Bertolazzi}, \citenamefont {Brivio},
  \citenamefont {Gao}, \citenamefont {Kis}, \citenamefont {Jena}, \citenamefont
  {Xing},\ and\ \citenamefont {Huang}}]{Shi2013}%
  \BibitemOpen
  \bibfield  {author} {\bibinfo {author} {\bibfnamefont {H.}~\bibnamefont
  {Shi}}, \bibinfo {author} {\bibfnamefont {R.}~\bibnamefont {Yan}}, \bibinfo
  {author} {\bibfnamefont {S.}~\bibnamefont {Bertolazzi}}, \bibinfo {author}
  {\bibfnamefont {J.}~\bibnamefont {Brivio}}, \bibinfo {author} {\bibfnamefont
  {B.}~\bibnamefont {Gao}}, \bibinfo {author} {\bibfnamefont {A.}~\bibnamefont
  {Kis}}, \bibinfo {author} {\bibfnamefont {D.}~\bibnamefont {Jena}}, \bibinfo
  {author} {\bibfnamefont {H.~G.}\ \bibnamefont {Xing}}, \ and\ \bibinfo
  {author} {\bibfnamefont {L.}~\bibnamefont {Huang}},\ }\href {\doibase
  10.1021/nn303973r} {\bibfield  {journal} {\bibinfo  {journal} {ACS Nano}\
  }\textbf {\bibinfo {volume} {7}},\ \bibinfo {pages} {1072} (\bibinfo {year}
  {2013})}\BibitemShut {NoStop}%
\bibitem [{\citenamefont {Sun}\ \emph {et~al.}(2014)\citenamefont {Sun},
  \citenamefont {Rao}, \citenamefont {Reider}, \citenamefont {Chen},
  \citenamefont {You}, \citenamefont {Brézin}, \citenamefont {Harutyunyan},\
  and\ \citenamefont {Heinz}}]{Sun2014}%
  \BibitemOpen
  \bibfield  {author} {\bibinfo {author} {\bibfnamefont {D.}~\bibnamefont
  {Sun}}, \bibinfo {author} {\bibfnamefont {Y.}~\bibnamefont {Rao}}, \bibinfo
  {author} {\bibfnamefont {G.~A.}\ \bibnamefont {Reider}}, \bibinfo {author}
  {\bibfnamefont {G.}~\bibnamefont {Chen}}, \bibinfo {author} {\bibfnamefont
  {Y.}~\bibnamefont {You}}, \bibinfo {author} {\bibfnamefont {L.}~\bibnamefont
  {Brézin}}, \bibinfo {author} {\bibfnamefont {A.~R.}\ \bibnamefont
  {Harutyunyan}}, \ and\ \bibinfo {author} {\bibfnamefont {T.~F.}\ \bibnamefont
  {Heinz}},\ }\href {\doibase 10.1021/nl5021975} {\bibfield  {journal}
  {\bibinfo  {journal} {Nano Lett.}\ }\textbf {\bibinfo {volume} {14}},\
  \bibinfo {pages} {5625} (\bibinfo {year} {2014})}\BibitemShut {NoStop}%
\bibitem [{\citenamefont {Siegert}(1939)}]{Siegert1936}%
  \BibitemOpen
  \bibfield  {author} {\bibinfo {author} {\bibfnamefont {A.~J.~F.}\
  \bibnamefont {Siegert}},\ }\href {\doibase 10.1103/PhysRev.56.750} {\bibfield
   {journal} {\bibinfo  {journal} {Phys. Rev.}\ }\textbf {\bibinfo {volume}
  {56}},\ \bibinfo {pages} {750} (\bibinfo {year} {1939})}\BibitemShut
  {NoStop}%
\bibitem [{\citenamefont {McCurdy}\ \emph {et~al.}(2004)\citenamefont
  {McCurdy}, \citenamefont {Baertschy},\ and\ \citenamefont
  {Rescigno}}]{McCurdy2004}%
  \BibitemOpen
  \bibfield  {author} {\bibinfo {author} {\bibfnamefont {C.~W.}\ \bibnamefont
  {McCurdy}}, \bibinfo {author} {\bibfnamefont {M.}~\bibnamefont {Baertschy}},
  \ and\ \bibinfo {author} {\bibfnamefont {T.~N.}\ \bibnamefont {Rescigno}},\
  }\href {http://stacks.iop.org/0953-4075/37/i=17/a=R01} {\bibfield  {journal}
  {\bibinfo  {journal} {J. Phys. B}\ }\textbf {\bibinfo {volume} {37}},\
  \bibinfo {pages} {R137} (\bibinfo {year} {2004})}\BibitemShut {NoStop}%
\bibitem [{\citenamefont {Simon}(1979)}]{Simon1979}%
  \BibitemOpen
  \bibfield  {author} {\bibinfo {author} {\bibfnamefont {B.}~\bibnamefont
  {Simon}},\ }\href {\doibase 10.1016/0375-9601(79)90165-8} {\bibfield
  {journal} {\bibinfo  {journal} {Phys. Lett.}\ }\textbf {\bibinfo {volume}
  {71}},\ \bibinfo {pages} {211 } (\bibinfo {year} {1979})}\BibitemShut
  {NoStop}%
\bibitem [{\citenamefont {McCurdy}\ \emph {et~al.}(1991)\citenamefont
  {McCurdy}, \citenamefont {Stroud},\ and\ \citenamefont
  {Wisinski}}]{McCurdy1991}%
  \BibitemOpen
  \bibfield  {author} {\bibinfo {author} {\bibfnamefont {C.~W.}\ \bibnamefont
  {McCurdy}}, \bibinfo {author} {\bibfnamefont {C.~K.}\ \bibnamefont {Stroud}},
  \ and\ \bibinfo {author} {\bibfnamefont {M.~K.}\ \bibnamefont {Wisinski}},\
  }\href {\doibase 10.1103/PhysRevA.43.5980} {\bibfield  {journal} {\bibinfo
  {journal} {Phys. Rev. A}\ }\textbf {\bibinfo {volume} {43}},\ \bibinfo
  {pages} {5980} (\bibinfo {year} {1991})}\BibitemShut {NoStop}%
\bibitem [{\citenamefont {Rescigno}\ and\ \citenamefont
  {McCurdy}(2000)}]{Rescigno2000}%
  \BibitemOpen
  \bibfield  {author} {\bibinfo {author} {\bibfnamefont {T.~N.}\ \bibnamefont
  {Rescigno}}\ and\ \bibinfo {author} {\bibfnamefont {C.~W.}\ \bibnamefont
  {McCurdy}},\ }\href {\doibase 10.1103/PhysRevA.62.032706} {\bibfield
  {journal} {\bibinfo  {journal} {Phys. Rev. A}\ }\textbf {\bibinfo {volume}
  {62}},\ \bibinfo {pages} {032706} (\bibinfo {year} {2000})}\BibitemShut
  {NoStop}%
\bibitem [{\citenamefont {Olsen}\ \emph {et~al.}(2016)\citenamefont {Olsen},
  \citenamefont {Latini}, \citenamefont {Rasmussen},\ and\ \citenamefont
  {Thygesen}}]{Olsen2016}%
  \BibitemOpen
  \bibfield  {author} {\bibinfo {author} {\bibfnamefont {T.}~\bibnamefont
  {Olsen}}, \bibinfo {author} {\bibfnamefont {S.}~\bibnamefont {Latini}},
  \bibinfo {author} {\bibfnamefont {F.}~\bibnamefont {Rasmussen}}, \ and\
  \bibinfo {author} {\bibfnamefont {K.~S.}\ \bibnamefont {Thygesen}},\ }\href
  {\doibase 10.1103/PhysRevLett.116.056401} {\bibfield  {journal} {\bibinfo
  {journal} {Phys. Rev. Lett.}\ }\textbf {\bibinfo {volume} {116}},\ \bibinfo
  {pages} {056401} (\bibinfo {year} {2016})}\BibitemShut {NoStop}%
\bibitem [{\citenamefont {Trinh}\ \emph {et~al.}(2013)\citenamefont {Trinh},
  \citenamefont {Tolstikhin}, \citenamefont {Madsen},\ and\ \citenamefont
  {Morishita}}]{Trinh2013}%
  \BibitemOpen
  \bibfield  {author} {\bibinfo {author} {\bibfnamefont {V.~H.}\ \bibnamefont
  {Trinh}}, \bibinfo {author} {\bibfnamefont {O.~I.}\ \bibnamefont
  {Tolstikhin}}, \bibinfo {author} {\bibfnamefont {L.~B.}\ \bibnamefont
  {Madsen}}, \ and\ \bibinfo {author} {\bibfnamefont {T.}~\bibnamefont
  {Morishita}},\ }\href {\doibase 10.1103/PhysRevA.87.043426} {\bibfield
  {journal} {\bibinfo  {journal} {Phys. Rev. A}\ }\textbf {\bibinfo {volume}
  {87}},\ \bibinfo {pages} {043426} (\bibinfo {year} {2013})}\BibitemShut
  {NoStop}%
\bibitem [{\citenamefont {Batishchev}\ \emph {et~al.}(2010)\citenamefont
  {Batishchev}, \citenamefont {Tolstikhin},\ and\ \citenamefont
  {Morishita}}]{Batishchev2010}%
  \BibitemOpen
  \bibfield  {author} {\bibinfo {author} {\bibfnamefont {P.~A.}\ \bibnamefont
  {Batishchev}}, \bibinfo {author} {\bibfnamefont {O.~I.}\ \bibnamefont
  {Tolstikhin}}, \ and\ \bibinfo {author} {\bibfnamefont {T.}~\bibnamefont
  {Morishita}},\ }\href {\doibase 10.1103/PhysRevA.82.023416} {\bibfield
  {journal} {\bibinfo  {journal} {Phys. Rev. A}\ }\textbf {\bibinfo {volume}
  {82}},\ \bibinfo {pages} {023416} (\bibinfo {year} {2010})}\BibitemShut
  {NoStop}%
\bibitem [{\citenamefont {Madsen}\ \emph {et~al.}(2012)\citenamefont {Madsen},
  \citenamefont {Tolstikhin},\ and\ \citenamefont {Morishita}}]{Madsen2012}%
  \BibitemOpen
  \bibfield  {author} {\bibinfo {author} {\bibfnamefont {L.~B.}\ \bibnamefont
  {Madsen}}, \bibinfo {author} {\bibfnamefont {O.~I.}\ \bibnamefont
  {Tolstikhin}}, \ and\ \bibinfo {author} {\bibfnamefont {T.}~\bibnamefont
  {Morishita}},\ }\href {\doibase 10.1103/PhysRevA.85.053404} {\bibfield
  {journal} {\bibinfo  {journal} {Phys. Rev. A}\ }\textbf {\bibinfo {volume}
  {85}},\ \bibinfo {pages} {053404} (\bibinfo {year} {2012})}\BibitemShut
  {NoStop}%
\bibitem [{\citenamefont {Madsen}\ \emph {et~al.}(2013)\citenamefont {Madsen},
  \citenamefont {Jensen}, \citenamefont {Tolstikhin},\ and\ \citenamefont
  {Morishita}}]{Madsen2013}%
  \BibitemOpen
  \bibfield  {author} {\bibinfo {author} {\bibfnamefont {L.~B.}\ \bibnamefont
  {Madsen}}, \bibinfo {author} {\bibfnamefont {F.}~\bibnamefont {Jensen}},
  \bibinfo {author} {\bibfnamefont {O.~I.}\ \bibnamefont {Tolstikhin}}, \ and\
  \bibinfo {author} {\bibfnamefont {T.}~\bibnamefont {Morishita}},\ }\href
  {\doibase 10.1103/PhysRevA.87.013406} {\bibfield  {journal} {\bibinfo
  {journal} {Phys. Rev. A}\ }\textbf {\bibinfo {volume} {87}},\ \bibinfo
  {pages} {013406} (\bibinfo {year} {2013})}\BibitemShut {NoStop}%
\bibitem [{\citenamefont {Yang}\ \emph {et~al.}(1991)\citenamefont {Yang},
  \citenamefont {Guo}, \citenamefont {Chan}, \citenamefont {Wong},\ and\
  \citenamefont {Ching}}]{Yang1991}%
  \BibitemOpen
  \bibfield  {author} {\bibinfo {author} {\bibfnamefont {X.~L.}\ \bibnamefont
  {Yang}}, \bibinfo {author} {\bibfnamefont {S.~H.}\ \bibnamefont {Guo}},
  \bibinfo {author} {\bibfnamefont {F.~T.}\ \bibnamefont {Chan}}, \bibinfo
  {author} {\bibfnamefont {K.~W.}\ \bibnamefont {Wong}}, \ and\ \bibinfo
  {author} {\bibfnamefont {W.~Y.}\ \bibnamefont {Ching}},\ }\href {\doibase
  10.1103/PhysRevA.43.1186} {\bibfield  {journal} {\bibinfo  {journal} {Phys.
  Rev. A}\ }\textbf {\bibinfo {volume} {43}},\ \bibinfo {pages} {1186}
  (\bibinfo {year} {1991})}\BibitemShut {NoStop}%
\bibitem [{\citenamefont {Dnestryan}\ and\ \citenamefont
  {Tolstikhin}(2016)}]{Dnestryan2016}%
  \BibitemOpen
  \bibfield  {author} {\bibinfo {author} {\bibfnamefont {A.~I.}\ \bibnamefont
  {Dnestryan}}\ and\ \bibinfo {author} {\bibfnamefont {O.~I.}\ \bibnamefont
  {Tolstikhin}},\ }\href {\doibase 10.1103/PhysRevA.93.033412} {\bibfield
  {journal} {\bibinfo  {journal} {Phys. Rev. A}\ }\textbf {\bibinfo {volume}
  {93}},\ \bibinfo {pages} {033412} (\bibinfo {year} {2016})}\BibitemShut
  {NoStop}%
\bibitem [{\citenamefont {Madsen}\ \emph {et~al.}(2017)\citenamefont {Madsen},
  \citenamefont {Jensen}, \citenamefont {Dnestryan},\ and\ \citenamefont
  {Tolstikhin}}]{Madsen2017}%
  \BibitemOpen
  \bibfield  {author} {\bibinfo {author} {\bibfnamefont {L.~B.}\ \bibnamefont
  {Madsen}}, \bibinfo {author} {\bibfnamefont {F.}~\bibnamefont {Jensen}},
  \bibinfo {author} {\bibfnamefont {A.~I.}\ \bibnamefont {Dnestryan}}, \ and\
  \bibinfo {author} {\bibfnamefont {O.~I.}\ \bibnamefont {Tolstikhin}},\ }\href
  {\doibase 10.1103/PhysRevA.96.013423} {\bibfield  {journal} {\bibinfo
  {journal} {Phys. Rev. A}\ }\textbf {\bibinfo {volume} {96}},\ \bibinfo
  {pages} {013423} (\bibinfo {year} {2017})}\BibitemShut {NoStop}%
\bibitem [{\citenamefont {Dnestryan}\ \emph {et~al.}(2018)\citenamefont
  {Dnestryan}, \citenamefont {Tolstikhin}, \citenamefont {Madsen},\ and\
  \citenamefont {Jensen}}]{Dnestryan2018}%
  \BibitemOpen
  \bibfield  {author} {\bibinfo {author} {\bibfnamefont {A.~I.}\ \bibnamefont
  {Dnestryan}}, \bibinfo {author} {\bibfnamefont {O.~I.}\ \bibnamefont
  {Tolstikhin}}, \bibinfo {author} {\bibfnamefont {L.~B.}\ \bibnamefont
  {Madsen}}, \ and\ \bibinfo {author} {\bibfnamefont {F.}~\bibnamefont
  {Jensen}},\ }\href {\doibase 10.1063/1.5046902} {\bibfield  {journal}
  {\bibinfo  {journal} {J. Chem. Phys.}\ }\textbf {\bibinfo {volume} {149}},\
  \bibinfo {pages} {164107} (\bibinfo {year} {2018})}\BibitemShut {NoStop}%
\bibitem [{\citenamefont {Mera}\ \emph {et~al.}(2015)\citenamefont {Mera},
  \citenamefont {Pedersen},\ and\ \citenamefont {Nikoli\ifmmode~\acute{c}\else
  \'{c}\fi{}}}]{Mera2015}%
  \BibitemOpen
  \bibfield  {author} {\bibinfo {author} {\bibfnamefont {H.}~\bibnamefont
  {Mera}}, \bibinfo {author} {\bibfnamefont {T.~G.}\ \bibnamefont {Pedersen}},
  \ and\ \bibinfo {author} {\bibfnamefont {B.~K.}\ \bibnamefont
  {Nikoli\ifmmode~\acute{c}\else \'{c}\fi{}}},\ }\href {\doibase
  10.1103/PhysRevLett.115.143001} {\bibfield  {journal} {\bibinfo  {journal}
  {Phys. Rev. Lett.}\ }\textbf {\bibinfo {volume} {115}},\ \bibinfo {pages}
  {143001} (\bibinfo {year} {2015})}\BibitemShut {NoStop}%
\bibitem [{\citenamefont {Dalgarno}\ \emph {et~al.}(1955)\citenamefont
  {Dalgarno}, \citenamefont {Lewis},\ and\ \citenamefont
  {Bates}}]{Dalgarno1955}%
  \BibitemOpen
  \bibfield  {author} {\bibinfo {author} {\bibfnamefont {A.}~\bibnamefont
  {Dalgarno}}, \bibinfo {author} {\bibfnamefont {J.~T.}\ \bibnamefont {Lewis}},
  \ and\ \bibinfo {author} {\bibfnamefont {D.~R.}\ \bibnamefont {Bates}},\
  }\href {\doibase 10.1098/rspa.1955.0246} {\bibfield  {journal} {\bibinfo
  {journal} {Proc. R. Soc. Lond. A}\ }\textbf {\bibinfo {volume} {233}},\
  \bibinfo {pages} {70} (\bibinfo {year} {1955})}\BibitemShut {NoStop}%
\bibitem [{\citenamefont {Pedersen}(2016)}]{Pedersen2016ExcitonStark}%
  \BibitemOpen
  \bibfield  {author} {\bibinfo {author} {\bibfnamefont {T.~G.}\ \bibnamefont
  {Pedersen}},\ }\href {\doibase 10.1103/PhysRevB.94.125424} {\bibfield
  {journal} {\bibinfo  {journal} {Phys. Rev. B}\ }\textbf {\bibinfo {volume}
  {94}},\ \bibinfo {pages} {125424} (\bibinfo {year} {2016})}\BibitemShut
  {NoStop}%
\bibitem [{\citenamefont {Scrinzi}(2010)}]{Scrinzi2010}%
  \BibitemOpen
  \bibfield  {author} {\bibinfo {author} {\bibfnamefont {A.}~\bibnamefont
  {Scrinzi}},\ }\href {\doibase 10.1103/PhysRevA.81.053845} {\bibfield
  {journal} {\bibinfo  {journal} {Phys. Rev. A}\ }\textbf {\bibinfo {volume}
  {81}},\ \bibinfo {pages} {053845} (\bibinfo {year} {2010})},\ \bibinfo {note}
  {{E}quations A1 and A2 in this paper contain misprints.
  $\left(C^{-1}\right)_{12}$ in Eq. A1 should be replaced by
  $\left(C^{-1}\right)_{21}$ and $\left(C^{-1}\right)_{21}$ in Eq. A2 should be
  replaced by $\left(C^{-1}\right)_{12}$.}\BibitemShut {Stop}%
\bibitem [{\citenamefont {Krylov}(2005)}]{Krylov2005}%
  \BibitemOpen
  \bibfield  {author} {\bibinfo {author} {\bibfnamefont {V.~I.}\ \bibnamefont
  {Krylov}},\ }\href@noop {} {\emph {\bibinfo {title} {Approximate Calculation
  of Integrals}}}\ (\bibinfo  {publisher} {Dover Publications},\ \bibinfo
  {year} {2005})\BibitemShut {NoStop}%
\end{thebibliography}%

\end{document}